\documentclass[lineno]{jfm}

\usepackage{gensymb}
\usepackage{subcaption}
\usepackage{graphicx}
\usepackage{newtxtext}
\usepackage{newtxmath}
\usepackage{natbib}
\usepackage{hyperref}
\hypersetup{
    colorlinks = true,
    urlcolor   = blue,
    citecolor  = black,
}

\newcommand{\RomanNumeralCaps}[1]
\linenumbers

\shorttitle{Displacement by hydrodynamic obstacle interaction in non-inertial flows}
\shortauthor{P. K. Das, X. Liu and S. Hilgenfeldt}

\title{Controlled particle displacement  by hydrodynamic obstacle interaction in non-inertial flows}

\author{Partha Kumar Das\aff{1},
  Xuchen Liu\aff{1},
 \and Sascha Hilgenfeldt\aff{1}\corresp{\email{sascha@illinois.edu}}}

\affiliation{\aff{1}Department of Mechanical Science and Engineering, University of Illinois Urbana-Champaign,
Urbana, IL 61801, USA}

\begin{document}

\maketitle

\begin{abstract}
Systematic deflection of microparticles off of initial streamlines is a fundamental task in microfluidics, aiming at applications including sorting, accumulation, or capture of the transported particles. 
In a large class of setups, including Deterministic Lateral Displacement and porous media filtering, particles in non-inertial (Stokes) flows are deflected by an array of obstacles. We show that net deflection of force-free particles passing an obstacle in Stokes flow is possible solely by hydrodynamic interactions if the flow and obstacle geometry break fore-aft symmetries. 
The net deflection is maximal for certain initial conditions and we analytically describe its scaling with particle size, obstacle shape, and flow geometry, confirmed by direct trajectory simulations. For realistic parameters, separation by particle size is comparable to what is found assuming contact (roughness) interactions.
Our approach also makes systematic predictions on when short-range attractive forces lead to particle capture or sticking. In separating hydrodynamic effects on particle motion strictly from contact interactions, we provide novel, rigorous guidelines for elementary microfluidic particle manipulation and filtering.
\end{abstract}

\begin{keywords}

\end{keywords}

\section{Introduction}\label{sec:introduction}

Microfluidic technologies have revolutionized particle manipulation, offering a diverse array of capabilities including transportation, separation, trapping, and enrichment \citep{pratt2011rare, sajeesh2014particle, lu2017particle}. Fundamentally, manipulation strategies force particles to cross streamlines so that they do not passively follow the flow. This essence of controlled manipulation, encompassing both passive (synthetic) or active (e.g.\ biological cells) particles, is implemented today in lab-on-a-chip processing as well as in the diagnosis of biological samples and biomanufacturing processes 
\citep{pamme2007continuous,ateya2008good,nilsson2009review,xuan2010particle,gossett2010label,puri2014particle}, drug discovery and delivery systems \citep{dittrich2006lab,kang2008microfluidics,nguyen2013design} or self-cleaning technologies \citep{callow2011trends,kirschner2012bio,nir2016bio}.  
Many techniques of precise control of suspended microparticles rely on the particle response to external forces including  
electrical \citep{xuan2019recent}, optical \citep{lenshof2010continuous,ashkin1986observation,grier2003revolution,chiou2005massively}, and magnetic techniques \citep{van2014integrated,crick1950physical,wang1993mechanotransduction,gosse2002magnetic}. However, not all particles are susceptible to these, prompting a continuous interest in manipulation strategies solely based on hydrodynamic forces. In practical applications, these approaches use customized flow geometries, such as channels or pillars 

\citep{lutz2006hydrodynamic,wiklund2012acoustofluidics,petit2012selective,tanyeri2011microfluidic,shenoy2016stokes,kumar2019orientation,chamolly2020irreversible}. In recent years, many such techniques leverage particle inertia \citep{di2007continuous,di2009inertial} particularly from oscillatory flow \citep{wang2011size,wang2012efficient,thameem2017fast,agarwal2018inertial} and have prompted an advance of theoretical frameworks beyond long-standing approaches \citep{maxey1983equation,gatignol1983faxen} to include new important effects \citep{rallabandi2021inertial,agarwal2021unrecognized,agarwal2021rectified,agarwal2024density}.

Inertial effects are not present in the Stokes limit $($Reynolds number $Re\rightarrow0)$, where hydrodynamic effects on particles are notoriously long-range and
have been described in fundamental detail \citep{happel1965low, brady1988stokesian, kim2013microhydrodynamics, pozrikidis1992boundary, pozrikidis2011introduction,rallabandi2017hydrodynamic}, while systematic modulation of particle trajectories in Stokes flow has not been the subject of a detailed study. 
Fundamentally, this is because of the instantaneity and time reversibility of Stokes flow that at first glance seems to preclude lasting particle displacements, despite the success of Deterministic Lateral Displacement (DLD) in sorting particles by size, forcing them onto different trajectories through the interaction with a forest of pillar obstacles at very low $Re$
\citep{huang2004continuous,kruger2014deformability,kabacaouglu2019sorting,hochstetter2020deterministic,lu2023label,aghilinejad2019transport,davis2006deterministic}. 

Size-based particle sorting in a typical DLD set-up is determined by whether a particle can cross separating streamlines (row-shift).  \citet{inglis2006critical,kulrattanarak2011analysis,pariset2017anticipating,cerbelli2012separation,DLDhydro1} heuristically developed critical length-scales for particle size above which lateral displacement by this row-shifting happens.
Modeling approaches for the interaction of particles with individual obstacles
generally appeal to contact forces or other strong short-range forces \citep{lin2002distance,dance2004collision} that break the time reversibility of the Stokes flow and the hydrodynamic forces following from it. In particular, \citet{frechette2009directional,balvin2009directional,bowman2013inertia} model the effects of a non-hydrodynamic repulsive force originating from surface roughness \citep{ekiel1999hydrodynamic} and its dependence on particle size. This approach has been informed by similar ideas
in the modeling of suspension rheology
\citep{da1996shear,metzger2010irreversibility,blanc2011experimental,pham2015particle,lemaire2023rheology}. 

Contact forces almost invariably contain an ad-hoc element \citep{ekiel1999hydrodynamic}, fundamentally because in strict Stokes flow, there is no surface contact in finite time \citep{brady1988stokesian,claeys1989lubrication,claeys1993suspensions}.
A recent study by \citet{li2024dynamics} shows that fibers passing over a triangular obstacle can show net displacements in Stokes transport flow even when not experiencing direct contact. The fibers are extended, non-spherical objects and their interactions with obstacles are modeled by effective forces  \citep{dance2004collision}, leaving open the question whether rigorously described hydrodynamic interactions can cause net displacement. Very recent work by \citet{liu2025principles} describes particle motion in {\em internal} (vortical) Stokes flow and finds that purely hydrodynamic interactions with confining channel walls  can indeed result in lasting displacements through repeated wall encounters.
Such vortex flows have, however, not been practically implemented yet for this purpose. The present work, by contrast, aims at a rigorous hydrodynamic description of particle motion in Stokes flow {\em external} to an obstacle (the elementary process of DLD), establishing the required geometry of flow and obstacle to effect a lasting displacement, as well as bounds on its magnitude.

\section{Hydrodynamic formalism of particle-wall interaction in Stokes flow}\label{sec.formalism}
Microparticles placed in a background Stokes flow near an interface experience forces at varying distances from the boundary.
Even at large distances (in bulk flow), particle trajectories deviate from the background flow because of Fax\'en's correction \citep{happel1965low}, but close to interfaces, the hydrodynamic interaction between boundaries is dominant.

Different situations can be addressed in the modeling of particles in Stokes flow \citep{brenner1961slow,goldman1967slow,goldman1967slow2}, particularly (i) the forces on a particle moving at a given speed, (ii) the forces on a particle held fixed in a certain location, or (iii) the motion of a force-free particle. The present work focuses on the third scenario, which is particularly relevant for density-matched particles (absent the effects of gravity) in microfluidic devices. 
Setting thus the total force on the particle to zero, we quantify the modification of the particle velocity due to the effects of nearby boundaries altering particle motion both in the wall-parallel and the wall-normal direction. 

A detailed analysis of the motion of force-free particles with wall-normal velocity corrections has been provided by \citet{rallabandi2017hydrodynamic}. Other work has described wall-parallel velocity corrections far from and near the wall \citep{ekiel2006accuracy, pasol2011motion}. In our very recent study on particle motion in internal Stokes flow \citep{liu2025principles}, we have constructed uniformly valid expressions for all particle-wall distances from this precursor work. Here, we extend the formalism to determine particle trajectories transported over an obstacle, representing a boundary whose orientation and curvature near the particle both change as the particle moves.

\subsection{Problem set-up}\label{sec prob setup}
Reflecting common situations in microfluidic setups with structures that span the entire height of a channel, we here describe 2-D Stokes flows $\boldsymbol{u}(\boldsymbol{x})$ in a Cartesian coordinate system $\boldsymbol{x}=(x,y)$. Although any 2-D Stokes flow around an obstacle is subject to the Oseen paradox \citep{proudman1957expansions} and does not exist as a consistent solution to arbitrary distance from the obstacle wall, a unique background Stokes flow exists in the vicinity of the obstacle, and the range and accuracy of that solution can be arbitrarily increased by lowering the Reynolds number. We place a spherical, force-free inertia-less particle in such a flow $\boldsymbol{u}(\boldsymbol{x})$ as sketched in figure~\ref{fig formalism}(a). 

We consider particle radii $a_p$ much smaller than the radius of curvature of the obstacle wall $R(\boldsymbol{x})$. The unit vector $\boldsymbol{e}_\perp(\boldsymbol{x_p})$ normal to the wall pointing towards the particle center $\boldsymbol{x}_p$ defines the closest distance $h(\boldsymbol{x_p})$ between particle center and boundary. 
Tracking the particle trajectory under the influence of the obstacle needs careful modeling of the corrections to both the wall-parallel and the wall-normal velocity components of the particle particularly in close proximity to the obstacle, where the effects are most prominent. In such close proximity, we have verified that effects of finite obstacle curvature (quantified in \cite{rallabandi2017hydrodynamic}) are small and do not alter the outcomes presented here (see Appendix~\ref{appen flat wall} for details), so that we restrict ourselves to the flat-wall approximation $(a_p/R\rightarrow 0)$ here.

Far from any walls, the motion of a spherical, neutrally buoyant particle is described by
\begin{equation}\label{vpfar}
    \boldsymbol{v}_{p,{far}} =\boldsymbol{u}(\boldsymbol{x}_p(t))+\boldsymbol{u}_{Faxen} (\boldsymbol{x}_p(t))\,,
\end{equation}
with the effect of streamline curvature
on the length scale of the particle size addressed by the Fax\'en correction, 
\begin{equation}\label{faxen}
    \boldsymbol{u}_{Faxen} (\boldsymbol{x})=\frac{a_p^2}{6}\nabla^2\boldsymbol{u}(\boldsymbol{x})\,.
\end{equation}
The presence of an obstacle imposes wall effects, and thus an additional velocity correction, $\boldsymbol{W}$. With no inertia, the particle equation of motion remains a first-order  dynamical system,
\begin{equation}\label{vp}
    \frac{d\boldsymbol{x}_{p}(t)}{dt}=\boldsymbol{v}_{p} (\boldsymbol{x}_p(t))=\boldsymbol{u}(\boldsymbol{x}_p(t))+\boldsymbol{u}_{Faxen} (\boldsymbol{x}_p(t))+\boldsymbol{W}(\boldsymbol{x}_p(t))\,.
\end{equation}

In the far field limit, $\boldsymbol{W}$ must decay to zero, leaving only the Fax\'en correction in effect. 
Decomposing the ambient velocity field 

$
\boldsymbol{u}=
u_{||} \boldsymbol{e}_{||} + u_\perp \boldsymbol{e}_{\perp}
$ 
in the wall-parallel and wall-normal directions, we write
\begin{equation}\label{vpparallel}
    \boldsymbol{v}_{p||} = \left(\boldsymbol{u} + \frac{a_p^2}{6}\nabla^2\boldsymbol{u}\right)\cdot\boldsymbol{e}_{||} + W_{||}\,,
\end{equation}
\begin{equation}\label{vpperp}
    \boldsymbol{v}_{p\perp} = \left(\boldsymbol{u} + \frac{a_p^2}{6}\nabla^2\boldsymbol{u}\right)\cdot\boldsymbol{e}_{\perp} + W_{\perp}\,,
\end{equation}
and quantify in the following the particle velocity corrections parallel to $(W_{||})$ and normal to $(W_{\perp})$ the wall.
While we largely quote results from previous work \citep{liu2025principles}, we point out where the present problem requires particular care and greater modeling effort.

\subsection{Wall-parallel corrections to the particle velocity}\label{sec wall parallel}
The components of $\boldsymbol{W}(\boldsymbol{x}_p)$ depend on the wall distance, conveniently described by the dimensionless parameter
\begin{equation}\label{eq Delta}
    \Delta=\frac{h(\boldsymbol{x}_p)-a_p}{a_p}\,,
\end{equation}
which is the surface-to-surface distance relative to the particle radius, cf.\ \citep{rallabandi2017hydrodynamic, thameem2017fast, agarwal2018inertial}. 

The wall-parallel velocity correction $W_{||}$ slows down the wall-parallel velocity of a force-free particle by a fraction $f(\Delta)$, so that

equation~\eqref{vpparallel} takes the form
\begin{equation}\label{eq vpparallelFull}
    v_{p||}(x_p,y_p) = \left[\left(1-f(\Delta)\right)\left(\boldsymbol{u} + \frac{a_p^2}{6}\nabla^2 \boldsymbol{u}\right)\cdot\boldsymbol{e}_{||} \right]_{\boldsymbol{x}_p}\,.
\end{equation}

A uniformly valid expression for $f(\Delta)$ was obtained in \citet{liu2025principles} by systematic asymptotic matching of results from \cite{goldman1967slow} for $\Delta\gg 1$ and  \cite{stephen1992characterization} for $\Delta\ll 1$ (including the lubrication limit), and is reproduced in Appendix~\ref{appen wall-parallel}.

\subsection{Wall-normal corrections to the particle velocity}\label{sec:wall normal}
Employing a quadratic expansion of the background flow about the spherical particle's center, a general expression for the normal component of the hydrodynamic force on a spherical, neutrally buoyant particle was obtained by \cite{rallabandi2017hydrodynamic}.
 
In the case of a force-free particle of interest here, this yields the wall-normal correction $W_\perp$ and thus the
particle wall-normal velocity 
\begin{equation} \label{eq vpPE}
    \begin{split}
    v_{p\perp}^{PE}(x_p,y_p)= &\left[\left\{\boldsymbol{u} + \frac{a_p^2}{6}\nabla^2 \boldsymbol{u} - a_p\frac{\mathcal{B}}{\mathcal{A}}(\boldsymbol{e_{\perp}}\cdot \nabla \boldsymbol{u}) + \frac{a_p^2}{2}\frac{\mathcal{C}}{\mathcal{A}}(\boldsymbol{e_{\perp}}\boldsymbol{e_{\perp}}:\nabla\nabla \boldsymbol{u}) + \right.\right. \\ & \left.\left. \frac{a_p^2}{2}\left(\frac{\mathcal{D}}{\mathcal{A}}-\frac{1}{3}\right)\nabla^2 \boldsymbol{u}\right\}\cdot\boldsymbol{e}_{\perp} \right]_{\boldsymbol{x}_p}\,.
    \end{split}
\end{equation}
The full analytical expressions for the $\Delta$-dependent functions $\mathcal{A}$, $\mathcal{B}$, $\mathcal{C}$ and $\mathcal{D}$ are provided in \citet{rallabandi2017hydrodynamic}. For arbitrarily large separations $(\Delta\rightarrow\infty)$, $\frac{\mathcal{B}}{\mathcal{A}}\to 0$, $\frac{\mathcal{C}}{\mathcal{A}}\to 0$, and $\frac{\mathcal{D}}{\mathcal{A}}\to 1/3$, so that $W_{\perp}\to 0$. We use the superscript $PE$ ("particle expansion") to emphasize that all background flow velocities and derivatives are evaluated at ${\bf x}_p$.

As the particle approaches the wall $(\Delta \ll 1)$ this particle expansion formalism becomes inaccurate -- note that \eqref{eq vpPE} is not guaranteed to vanish when the particle touches the wall $(\Delta=0)$. If instead we employ Taylor expansion of the background flow field $\boldsymbol{u}(\boldsymbol{x})$ around the point on the wall nearest to the particle, 
\begin{equation}\label{eq xw}
\boldsymbol{x}_w
=\boldsymbol{x}_p
-a_p\boldsymbol{e}_{\perp}(\boldsymbol{x}_p)\,,
\end{equation}
as suggested by \citet{rallabandi2017hydrodynamic}, the no-penetration condition is enforced ("wall expansion"). The resulting particle velocity is linear in $\Delta$ to leading order,
\begin{equation}\label{eq wall xpnsn}
    v_{p\perp}^{WE} = 1.6147 
 a_p^2\kappa(\boldsymbol{x}_p)\Delta+O(\Delta^2)\,.
\end{equation}
Here $\kappa(\boldsymbol{x}_p)=\partial_{\perp}^2u_{\perp}(\boldsymbol{x}_w)$ is the background flow curvature at the wall point $\boldsymbol{x}_w$ from \eqref{eq xw}.
 
Although \eqref{eq wall xpnsn} accurately determines the particle wall-normal velocity in the near-wall limit $(\Delta\ll1)$, we find that it does not smoothly transition to \eqref{eq vpPE} when $\Delta\sim1$. We therefore generalize and refine the wall-expansion procedure in this intermediate region by constructing second-order expansions of $\boldsymbol{u}(\boldsymbol{x})$ around variable expansion points $\boldsymbol{x}_E=\boldsymbol{x}_E(\boldsymbol{x}_p)$, i.e., 
\begin{equation} \label{upVE}
    \begin{split}
    \boldsymbol{u}^{VE}(\boldsymbol{x}) = \boldsymbol{u}|_{\boldsymbol{x}=\boldsymbol{x}_E} + (\boldsymbol{x}-\boldsymbol{x}_E)\cdot(\nabla \boldsymbol{u})_{\boldsymbol{x}=\boldsymbol{x}_E} + \frac{1}{2}(\boldsymbol{x}-\boldsymbol{x}_E)(\boldsymbol{x}-\boldsymbol{x}_E):(\nabla \nabla \boldsymbol{u})_{\boldsymbol{x}=\boldsymbol{x}_E}\,.
    \end{split}
\end{equation}
Replacing $\boldsymbol{u}$ by $\boldsymbol{u}^{VE}$ in \eqref{eq vpPE} gives the generalized expression
\begin{equation} \label{eq vpVE}
    \begin{split}
        v_{p\perp}^{VE}(x_p,y_p) = &\left[\left\{\boldsymbol{u}^{VE} + \frac{a_p^2}{6}\nabla^2 \boldsymbol{u}^{VE} - a_p\frac{\mathcal{B}}{\mathcal{A}}(\boldsymbol{e_{\perp}}\cdot \nabla \boldsymbol{u}^{VE}) + \frac{a_p^2}{2}\frac{\mathcal{C}}{\mathcal{A}}(\boldsymbol{e_{\perp}}\boldsymbol{e_{\perp}}:\nabla\nabla \boldsymbol{u}^{VE}) + \right.\right. \\ & \left.\left. \frac{a_p^2}{2}\left(\frac{\mathcal{D}}{\mathcal{A}}-\frac{1}{3}\right)\nabla^2 \boldsymbol{u}^{VE}\right\}\cdot\boldsymbol{e}_{\perp} \right]_{\boldsymbol{x}_p}\,,
    \end{split}
\end{equation}
where the derivatives and resistance coefficients are still evaluated at the particle center. 

The functional form of the expansion point 
$\boldsymbol{x}_E(\boldsymbol{x}_p)$
is constructed for \eqref{eq vpVE} to obey the WE and PE limits, i.e 
$\boldsymbol{x}_E\to \boldsymbol{x}_w$ for touching particles $(\Delta \to 0)$ and 
$\boldsymbol{x}_E = \boldsymbol{x}_p$ for all $\Delta\geq \Delta_E$.
For simplicity, we choose the linear relation 
\begin{equation}\label{eq xE}
    \boldsymbol{x}_E=\boldsymbol{x}_p-a_p\left(1-\frac{\Delta}{\Delta_E}\right)\boldsymbol{e}_{\perp}\,,
\end{equation}
for $\Delta\leq \Delta_E$ and $\boldsymbol{x}_E=\boldsymbol{x}_p$ otherwise.
 
\begin{figure}
    \centering
\includegraphics[width=\textwidth]{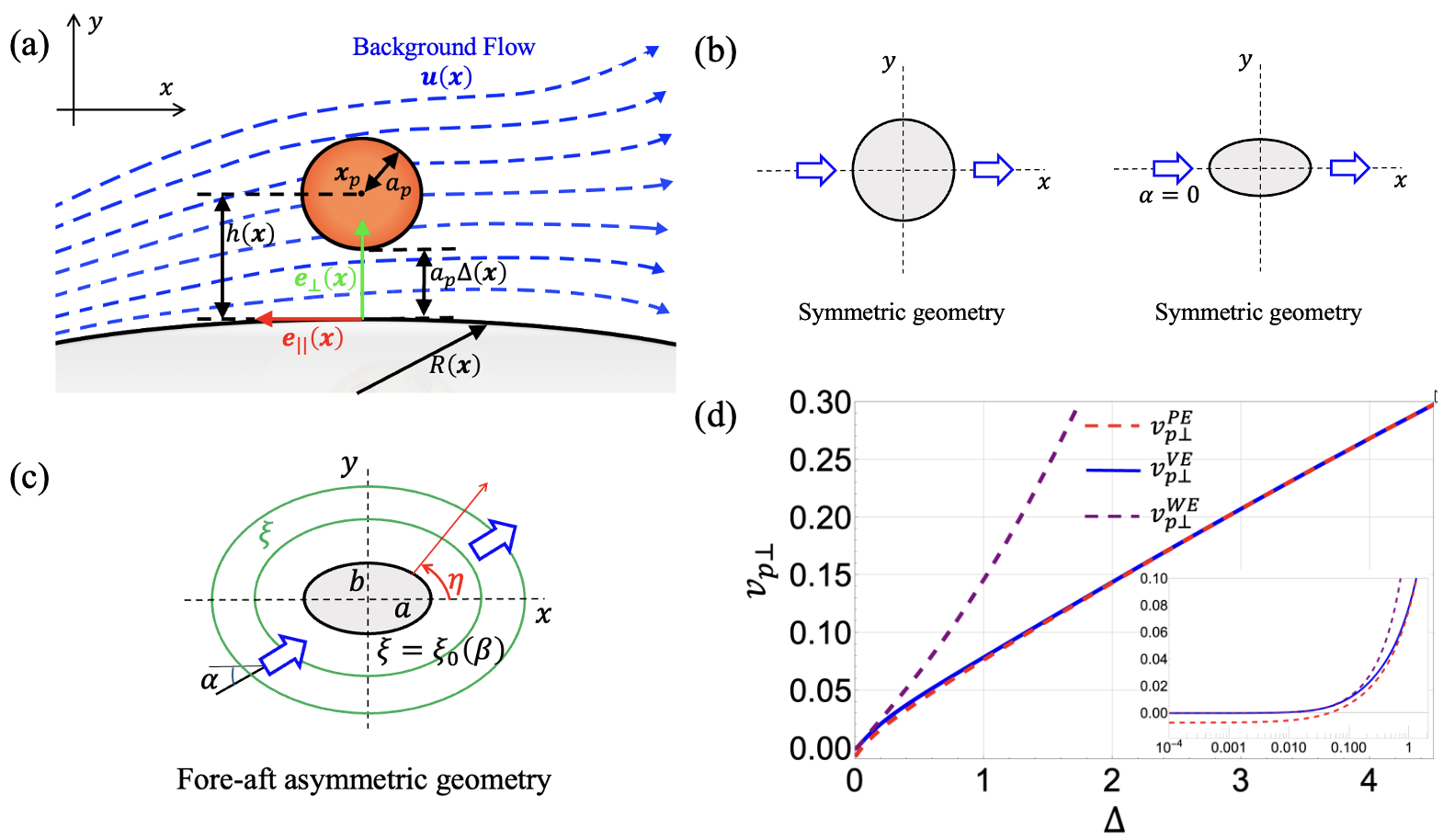}    \caption{(a) Schematic of a small spherical particle of radius $a_p$ located at $\boldsymbol{x}_p=(x_p,y_p)$ in the vicinity of an obstacle boundary of radius of curvature $R(\boldsymbol{x}$). The particle is immersed in a density-matched Stokes background flow $\boldsymbol{u}(\boldsymbol{x})$. 
(b) sketches of fore-aft symmetric obstacle-flow geometries with a circular cylinder or a symmetrically placed elliptic cylinder.  This fore-aft symmetry breaks in (c), where the flow has non-trivial angle of attack $\alpha$. Also sketched is the elliptic coordinate system $(\xi,\eta)$. (d) An example of wall-normal particle velocity as a function of the gap coordinate $\Delta$ at an angular position $\eta=20^\degree$ 
using variable expansion modeling (cf. \eqref{eq vpVE}) with $\Delta_E=3$ showing a smooth transition from particle expansion for large $\Delta$ to the wall expansion (inset) for small $\Delta$. 
} \label{fig formalism}
\end{figure}

Figure~\ref{fig formalism}(b) shows an example of the smooth interpolation between $v_{p\perp}^{WE}$ and $v_{p\perp}^{PE}$ for 
$\Delta_E=3$. We adopt this choice for subsequent sections, while carefully checking that the results are robust against other ${\cal O}(1)$ choices of $\Delta_E$ (appendix~\ref{appen modeling}).

Using equations \eqref{eq vpparallelFull} and \eqref{eq vpVE} in the dynamical system \eqref{vp} constitutes our formalism for computing particle motion in the presence of wall effects for arbitrary Stokes background flow.

\section{Symmetry-breaking inertialess transport around an obstacle}\label{sec.symmetry breaking}
We apply the modeling detailed above to particles transported towards and past an obstacle. 
If the particle’s approach to and departure from the interface on its trajectory occur in a symmetric fashion, the gradients normal to the wall cancel out during approach and departure, so that no net displacement (relative to the streamline the particle starts on) will be observed. 
This is clearly the case for a single circular cylinder obstacle (figure~\ref{fig formalism}b). 
However, if the combined geometry of the boundary and the  flow field break this symmetry, net displacement and thus meaningful manipulation of particle transport is possible. We investigate here the case of an elliptic cylinder obstacle with aspect ratio $\beta=\frac{b}{a}$ (with $a$ and $b$ the major and minor axis, respectively) placed in a uniform Stokes flow $U$ whose direction makes an angle $\alpha$ with the major axis. This situation is shown in figure~\ref{fig formalism}(c) together with associated cartesian and elliptic coordinate systems. Note that symmetry arguments again preclude net displacement if $\alpha=0$ or $\alpha=\pi/2$ (figure~\ref{fig formalism}b).

We use the single elliptic obstacle as the simplest case study of fore-aft symmetry breaking, which in 
traditional DLD setups is achieved by inclining a forest of circular pillars relative to the uniform stream. 
The Stokes background flow around the elliptic cylinder is known analytically as a solution to the 
biharmonic equation
    $\nabla^4\tilde{\psi}_B=0$ 
for the stream function $\tilde{\psi}_B$.
We choose the semimajor axis $a$ and the uniform flow speed $U$ as our length and velocity scales, so that $\tilde{\psi}_B$ is non-dimensionalized by $aU$. 

\begin{figure}  
    \centering
\includegraphics[width=\textwidth]{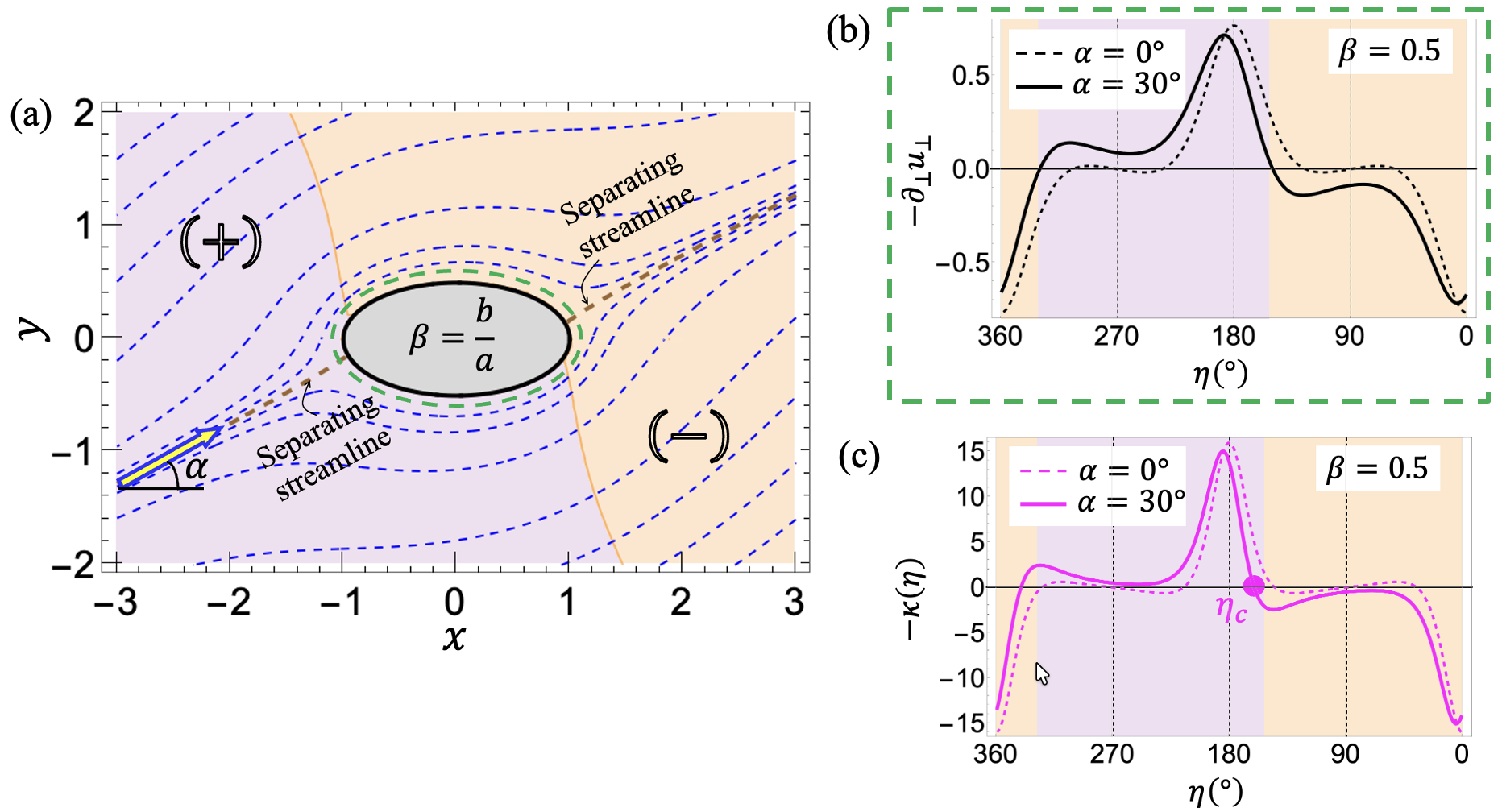}    \caption{(a) Fore-aft asymmetric Stokes flow (dashed blue streamline contours from \eqref{eq psiB}) impacting on an elliptic obstacle ($\beta=1/2$) under an inclination $\alpha=30\degree$. 
$W_\perp$ is positive in the purple shaded zone and is negative in the orange shaded zone indicating particle repulsion from the obstacle and attraction towards the obstacle, respectively. 
The separating streamlines $\psi=0$ are indicated in dashed brown. 
(b) The sign changes of the quantity $\partial_{\perp} u_{\perp}$ reflect, to  leading order, those of $W_{\perp}$ (shading). This quantity is measured along the green dashed line in (a), locations at a distance $\Delta = 1$, here for $a_p=0.1$, where $\partial_{\perp} u_{\perp}$ dominates the corrections of particle motion. (c) The background flow curvature $-\kappa=-\partial^2_{\perp}u_{\perp}$ evaluated at the obstacle wall determines normal particle motion in the wall expansion model \eqref{eq wall xpnsn} for $\Delta\ll 1$. The upstream zero of this quantity, $\eta_c$, indicates a point of closest approach 
to the obstacle as discussed in section \S\ref{sec closest approach}.
}  \label{fig flow}
\end{figure}

The Oseen paradox restricts this solution to an inner region in the vicinity of the obstacle surface.
It is implicit in the work of
\cite{berry1923steady} and was explicitly derived
by \cite{shintani1983low} in elliptic coordinates $(\xi,\eta)$ as the infinite sum $\tilde{\psi}_B=\sum_{n=1}^\infty\tilde{\psi}_n$ with
\begin{equation}\label{eq flow}
    \begin{split}
        \tilde{\psi}_n= \frac{1}{(\ln Re)^n}[&\Lambda_n \{(\xi-\xi_0)\cosh {\xi} + \sinh \xi_0 \cosh\xi_0\cosh\xi - \cosh^2\xi_0 \sinh\xi \}\cos\eta - \\& \Omega_n \{(\xi-\xi_0)\sinh\xi -\sinh\xi_0\cosh\xi_0 \sinh\xi + \sinh^2 \xi_0 \cosh\xi\}\sin\eta]\,.
    \end{split}
\end{equation}
Here $x=\sqrt{1-\beta^2} \cosh\xi\cos\eta$ and  $y=\sqrt{1-\beta^2} \sinh\xi\sin\eta$ relate cartesian and elliptic coordinates, and $\xi=\xi_0=\frac{1}{2} \ln\frac{1+\beta}{1-\beta}$ defines the elliptic cylinder surface. The prefactors $\Lambda_n$ and $\Omega_n$ are functions of $\alpha$ and $\beta$ obtained from asymptotic matching \citep{proudman1957expansions,kaplun1957low}. The analytical form of  \eqref{eq flow}, by construction, does not contain fluid inertia except an implicit dependence on Reynolds number $Re$ which defines the Oseen distance
\citep{proudman1957expansions},
which can be made arbitrarily large by choosing a very small $Re$. Since our goal is to investigate the wall effect on particle trajectory and effective particle-wall interaction always occurs in the proximity of the obstacle, this flow description reflects reality to a well-defined degree of accuracy. In the limit $Re\rightarrow0$, the $n=1$ term of \eqref{eq flow} is dominant, and we scale out the $Re$ dependence by defining the rescaled background flow as
\begin{equation}\label{eq psiB}
\begin{split}
    \psi_B=(-\ln Re)\tilde{\psi}_{n=1}= &\Lambda_1 \{(\xi-\xi_0)\cosh {\xi} + \sinh \xi_0 \cosh\xi_0\cosh\xi - \cosh^2\xi_0 \sinh\xi \}\cos\eta - \\& \Omega_1 \{(\xi-\xi_0)\sinh\xi -\sinh\xi_0\cosh\xi_0 \sinh\xi + \sinh^2 \xi_0 \cosh\xi\}\sin\eta  \,,
\end{split}
\end{equation}
with $(\Lambda_1,\Omega_1)
    =\sqrt{1-\beta^2}
    (\sin\alpha, \cos\alpha)$. 
This analytical expression agrees with the description of Stokes flow over an inclined elliptic fiber by  \cite{raynor2002flow}. Figure~\ref{fig flow}(a) shows streamline contours $(\psi_B)$ for an angle of attack  $\alpha = 30\degree$ and obstacle aspect ratio $\beta=0.5$.  Lastly, in order to isolate and explicitly quantify the wall effect on particle motion only, we use as our reference streamfunction the sum of $\psi_B$ and the Fax\'en streamfunction $\psi_{Faxen}=\frac{a_p^2}{6}\nabla^2\psi_B$, i.e.,
\begin{equation}\label{eq faxen flow}
    \begin{split}
        \psi = \psi_B+\psi_{Faxen}\,.
    \end{split}
\end{equation}
We have confirmed (see appendix \ref{appen bulk Faxen}) that particle displacement effects from the Fax\'en correction are negligible compared to those from particle-wall interaction at small $\Delta$, while they are absolutely small when $\Delta\gg 1$ and thus do not affect our findings on net displacement resulting from particle-obstacle encounters. 

\section{Results and discussion}\label{sec result}
The hydrodynamic formalism of section \S \ref{sec.formalism} is now applied to compute the particle trajectory of transport around an elliptic obstacle in the inertialess flow $\psi$ from \eqref{eq faxen flow}. The equations of motion \eqref{eq vpparallelFull}, \eqref{eq vpVE} use the background flow velocity $\boldsymbol{u}$ derived from $\psi$ in \eqref{eq psiB} via $u_\eta(\xi,\eta)=-g\partial_\xi\psi$, $u_\xi(\xi,\eta)=g\partial_\eta\psi$ where $g=g(\xi,\eta)=[(1-\beta^2)(\cosh^2\xi-\cos^2\eta)]^{-1/2}$  is the scale factor of the elliptic coordinate system \citep{shintani1983low,raynor2002flow}. For numerical computations, we transform all equations into cartesian reference coordinates, while for some analytical arguments, we will use elliptic coordinates directly. 


Qualitatively, the particle is transported towards the obstacle surface on the upstream side and away from it on the downstream side. If the initial position of the particle leads to a very close approach to the obstacle wall (or even a hypothetical overlap with passive transport), $W_{\perp}$ will act to repel the particle from the wall on the upstream side. Conversely, $W_{\perp}$ represents an attraction downstream. Figure~\ref{fig flow}(a) confirms these observations and shows that for non-trivial $\alpha,\beta$ the zones of repulsion and attraction are strongly asymmetric.

We will be most interested in particle trajectories that follow the obstacle outline closely, i.e., where $\Delta$ is not large. It can be verified that at $\Delta\sim 1$, the magnitude of $W_{\perp}$ is dominated by the first normal derivative term $-a_p\frac{\mathcal{B}}{\mathcal{A}}\partial_{\perp}u_{\perp}$ in \eqref{eq vpPE}. Indeed, figure~\ref{fig flow}(b) confirms that the sign of $-\partial_{\perp}u_{\perp}$ determines total attraction $(-)$ or repulsion $(+)$ to good accuracy for points at a distance of $h=2a_p$
from the wall ($\Delta=1$).


In the limit $\Delta \to 0$, the net effect of $W_{\perp}$ on the particle motion is determined to leading order of $\Delta$ by the wall flow curvature $\kappa=\partial_{\perp}^2 u_{\perp}$,
cf.\ \eqref{eq wall xpnsn}. 
Figure~\ref{fig flow}(c) demonstrates that $\kappa$ changes sign in very similar angular positions as the full $W_{\perp}$. We focus on information about $W_\perp$ here, as its effect dominates particle streamline crossing, while $W_\parallel$ mainly serves to slow particles down on their path.

We will illustrate particle trajectories and displacements in the following with particles that travel from left to right "above" the obstacle, where the repulsion region $(+)$, encountered first, is shorter than the attraction region $(-)$. All behavior of particles traveling  "below" the obstacle can be inferred by symmetry as discussed later.
The cases "above" and "below" have two separating streamlines ($\psi=0$) as boundaries,  representing a remaining flow symmetry: because of time reversibility of the Stokes flow (and the resulting hydrodynamic effects, which are linear in the Stokes flow), no particle can cross over these separating streamlines, which intersect the obstacle at the angular coordinate  $\eta^{sep}=\arctan(\beta\tan\alpha)$ downstream and $\pi+\eta^{sep}$ upstream, respectively. 


\subsection{Particle trajectories with net displacement}\label{sec trajectory}

The equation of motion of the particle is solved numerically to obtain the trajectory $\boldsymbol{x_p}(t)$. We define a complete journey of a particle from left to right as  starting from an initial $x=x_i$ position and ending at $x_f=-x_i$ position. 
It is  important to note that: \\
(i) The inherent linearity in Stokes flow causes the wall effects $\boldsymbol{W}$ to be fully time-reversible, meaning that particles will come back exactly to their starting points in reversing the flow. \\
(ii) Stokes flow does not possess any memory from inertia. Thus, the effect on the particle trajectory by $\boldsymbol{W}$ is instantaneous, and an initial particle position determines the full trajectory.\\
(iii) Because of the anti-symmetry of the flow about the separating streamlines, a trajectory transported above the obstacle from an initial stream function value $\psi=\psi_i$ is equivalent to a trajectory starting at $\psi=-\psi_i$ that is transported below in the opposite direction.\\

\begin{figure}
  \centering
  \begin{subfigure}[t]{\textwidth}
    \centering
    \includegraphics[width=\linewidth]{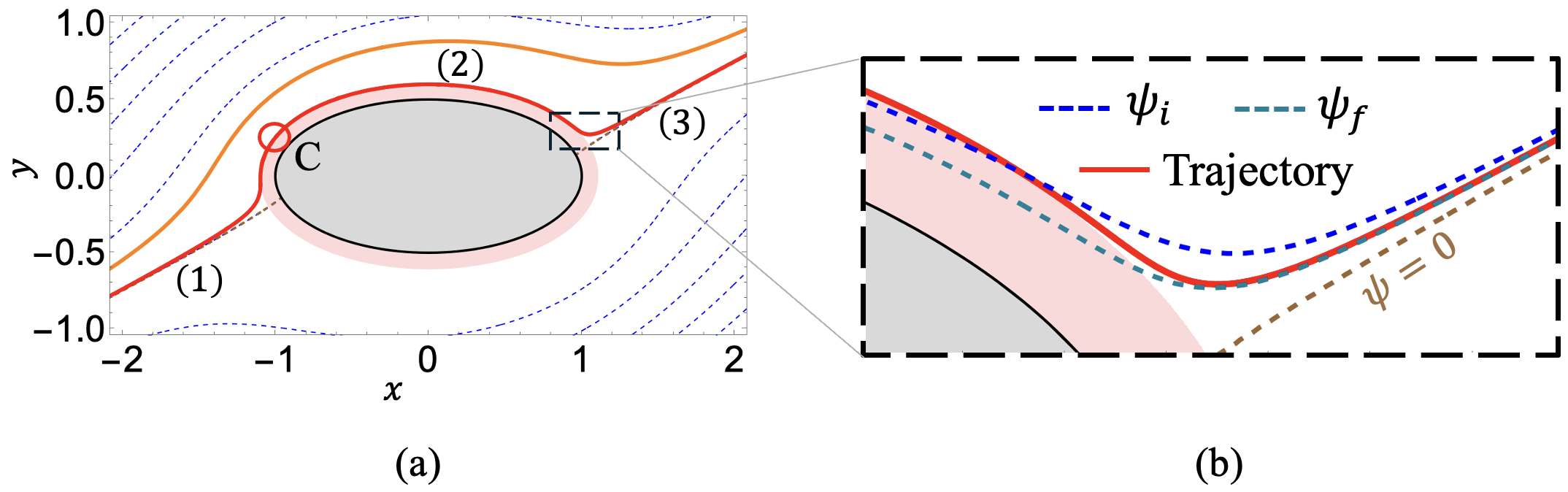} 
  \end{subfigure}

  \vspace{2em}

 \begin{subfigure}[t]{\textwidth}
    \centering
    \includegraphics[width=\linewidth]{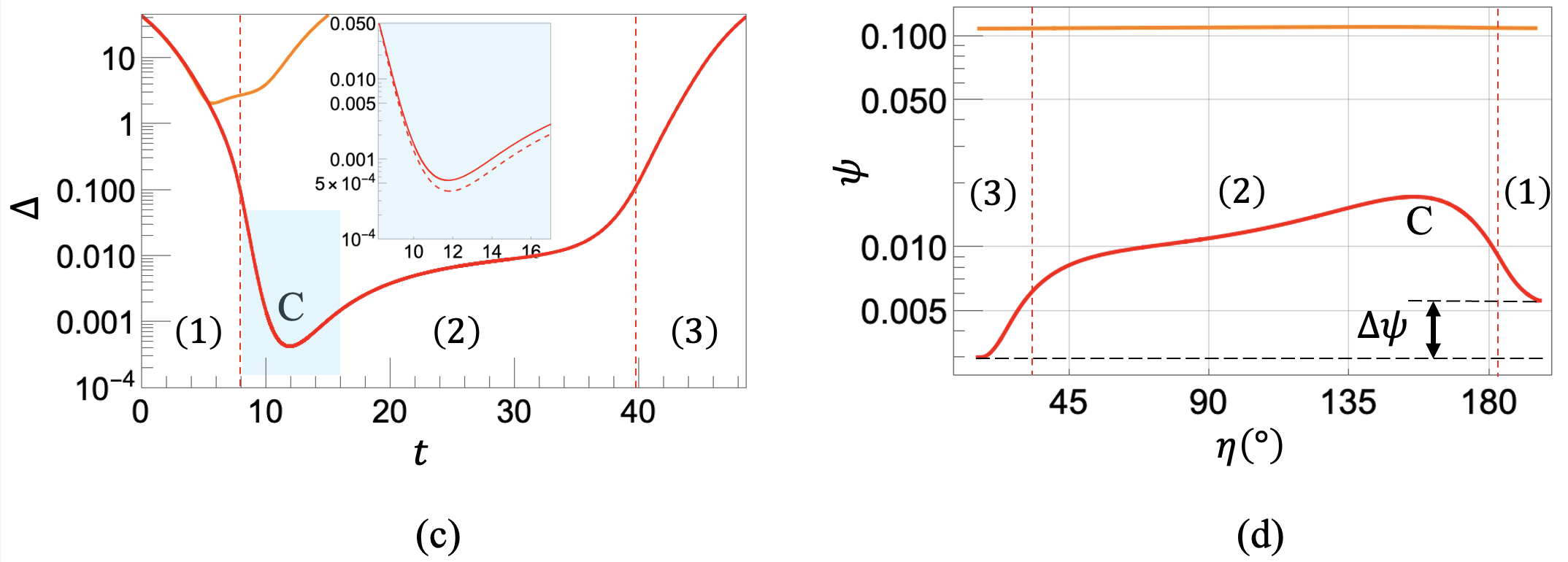}
  \end{subfigure}
  \caption{(a) Two computed trajectories (orange and red) of an $a_p=0.1$ particle above an obstacle ($\beta=0.5$) and $\alpha=30\degree$ inclined Stokes flow ($\psi$ isolines in dashed blue). Red shading indicates the exclusion zone for the hard-sphere particle center. (b) Magnified downstream view shows that the final streamline the particle asymptotes to is lower than its initial streamline ($\psi_f<\psi_i$). 
(c) Particle-wall gap $\Delta$ as a function of travel time. The orange trajectory stays far from the wall ($\Delta>1$), while the red trajectory remains very close ($\Delta\ll1$). The inset shows good agreement around the minimum with computations using the wall expansion model \eqref{eq wall xpnsn} (red dashed line).
(d) Local value of stream function on the trajectories as a function of angular position $\eta$. The orange trajectory remains essentially undeflected, while the red trajectory shows the effects of strong deflection away (1-2) and towards (2-3) the obstacle. 
Asymmetry of the flow ensures a net change in streamfunction $\Delta\psi=|\psi_i-\psi_f|$. 
}
  \label{fig trajectory}
\end{figure}

For empirical computations, unless otherwise stated, we choose $a_p=0.1$ and track the particle center positions to get  pathlines.  Two representative trajectories are shown in figure~\ref{fig trajectory}(a): one (in orange) never gets very close to the obstacle ($\Delta>1$ throughout), while the other particle (in red) spends the majority of time creeping around the obstacle at very small gap values ($\Delta\ll 1$), a trajectory called a "dive" in porous media literature \cite{miele2025flow}. 
Figure~\ref{fig trajectory}(c) shows the dynamics of the gap $\Delta$. The inset magnifies the closest approach of the dive trajectory towards the wall (position C), and verifies that the dynamics there is closely approximated by the leading order wall expansion model \eqref{eq wall xpnsn}. 


With an initial condition reasonably far  from the obstacle (a situation common in microfluidics devices, we choose $x_i=-5$), at first the particle moves along its initial streamline $\psi=\psi_i$ (stage (1) in figure~\ref{fig trajectory}). When closer to the obstacle, the wall-normal correction counteracts the background flow still pointing towards the obstacle. At position C, the normal flow velocity $u_{\perp}$ and the wall effect $W_{\perp}$ cancel, so that
 the particle normal velocity $v_{p\perp}=0$ and the particle-obstacle gap $\Delta$ reaches its minimum $\Delta_{min}$ (figure~\ref{fig trajectory}c).

 
Proceeding further in the dive stage (2), the wall effect $W_\perp$ becomes negative (cf.\ figures~\ref{fig flow}(b) and ~\ref{fig flow}(c)) and thus pushes the particle towards the obstacle while the background flow causes transport away from the wall. This continues until the particle ends its travel downstream on a well-defined final streamline $\psi=\psi_f$ (stage (3) in figure~\ref{fig trajectory}). 
For the situation depicted here ($\alpha<\pi/2$), we see that the 
phase of inward pull is of greater extent than that of outward repulsion, and that indeed there is a net downward displacement of the particle  ($\psi_f<\psi_i$, see figure~\ref{fig trajectory}b). 


It is convenient to quantify displacements by changes of the instantaneous stream function value of the particle position.
Figure~\ref{fig trajectory}(d) shows $\psi$ as a function of angular elliptic coordinate $\eta$. While the trajectory far from the obstacle (orange) shows negligible changes, the dive trajectory (red, very small $\psi_i$) registers a rapid increase of $\psi$ in the dive phase up to C, and then a gradual decrease to a lasting displacement with well-defined
$\Delta \psi = |\psi_f-\psi_i|$, which quantifies the strength of net deflection due to the symmetry-broken flow geometry. 

At first glance, this implies larger $\Delta\psi$ as $\psi_i$ decreases. However, when the initial position approaches the separating streamline $\psi_i\rightarrow 0$, this tendency cannot continue: particles cannot cross the separating streamline for symmetry reasons (see above), so $\psi_f\to 0$ is required and also the (downward) deflection must vanish, $\Delta\psi\to 0$. 
This reasoning guarantees a characteristic  maximum value of net displacement $\Delta\psi$ for a particular $\psi_i$.
In the following subsections we systematically describe this surprising effect both numerically and analytically. 

\subsection{Analysis of net displacements}
\subsubsection{Integrating the dynamical system} \label{sec empirical}
Numerical integration of the particle trajectory was performed as described in section \S\ref{sec.formalism} for varying initial conditions. We fix $x_i=-5$ and vary $y_i$ in $\psi_i=\psi(x_i,y_i)$. The final stream function value is then assessed as $\psi_f=\psi(-x_i,y_f)$ when the particle's $x$-position reaches $x_p=-x_i$. We choose $|x_i|$ large enough that it does not influence the result, i.e., changes in $\psi$ for particle positions $x<x_i$ or $x>x_f$ are negligible.

\begin{figure}
    \centering
\includegraphics[width=\textwidth]{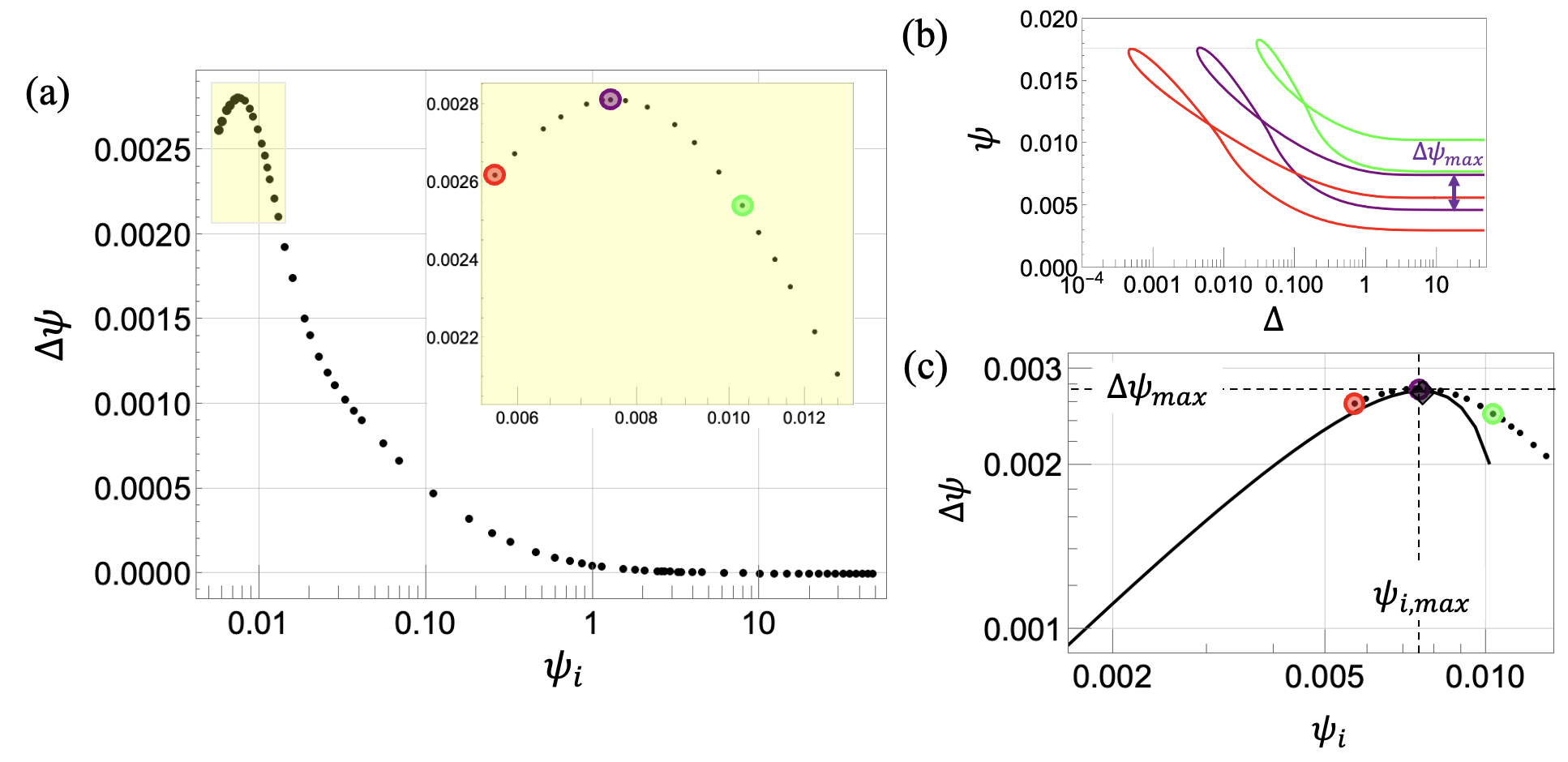}    \caption{Particle net displacement results for $(\alpha,\beta,a_p) = (30\degree,0.5,0.1)$. (a) Plot of displacement $\Delta\psi$ for particle trajectories released from different initial streamlines $\psi_i$. The inset magnifies the region around the maximum value $\Delta\psi_{max}$. (b) Variation of $\psi$ with $\Delta$ along the trajectories corresponding to the colored points in (a), including the $\Delta\psi_{max}$ trajectory in purple. (c) The solid line depicts  $\Delta\psi(\psi_i)$ from the analytical model equation~\eqref{eq etafinal etaini} for small $\psi_i$. Both the magnitude of $\Delta\psi_{max}$ and its location $\psi_{i,max}$ are in good agreement with the empirical calculations. 
}\label{fig deflection}
\end{figure}

Figure~\ref{fig deflection}(a) shows the variation of the deflection measure $\Delta\psi$ with $\psi_i$ for $\beta=1/2, \alpha=30\degree$.
Particles that start far away from the separating streamline ($\psi_i \gtrsim 1$) experience minimal wall interaction throughout and $\Delta\psi$ remains nearly zero. Particles accumulate meaningful wall effect, and thus net displacement, as they encounter the obstacle in closer proximity ($\Delta<1$) when starting at smaller $\psi_i$. The computations confirm the general argument above, showing a prominent maximum in $\Delta\psi$ at a small value of $\psi_i$.
The inset highlights three initial condition before (green), at (purple), and beyond (red) the maximum.
In figure~\ref{fig deflection}(b) we show the  changes in $\psi$ and $\Delta$ on these three particle trajectories. 


The dynamical system is solved in Mathematica with controlled accuracy, which for initial positions  $\psi_i \lesssim 0.005$ 
becomes computationally too demanding. The maximum in $\Delta \psi$ is, however, well resolved and we will show below that we can understand the behavior for $\psi_i\to 0$ analytically. The value of the maximum, $\Delta\psi_{max}$, and the corresponding initial position $\psi_{i,max}$ are main results of the current work that characterize the maximum net deflection that can be expected from a given symmetry-broken obstacle.



We stress that the stream function changes quantified here are with respect to the 
reference flow $\psi$ from \eqref{eq faxen flow} and thus explicitly quantify wall effects only. The Fax\'en term in the reference flow describes the bulk flow curvature effects on particle trajectories (relative to passive transport in the background Stokes flow). We confirm 
in appendix \ref{appen bulk Faxen} that deflection by the Fax\'en term in the absence of the wall is at least one order of magnitude smaller than the  wall effect, so that we isolate the main physics here. 

\subsubsection{Analytical formalism}\label{analytical}
In this section, we develop a method of predicting $\Delta\psi_{max}$, particularly in the limit $\psi_i\to 0$, which proves computationally challenging. We shall see that only information directly derived from the background flow field is needed.


All trajectories for $\psi_i\lesssim \psi_{i,max}$ are of the "red" type in figure~\ref{fig trajectory}, i.e., they (1) approach the obstacle in close proximity of the upstream separating streamline (the angular particle position is $\eta_i\approx \pi+\eta^{sep}$), (2)
accumulate meaningful deflection while moving along the obstacle surface maintaining very small gaps $\Delta\ll 1$ ("dives",  \citep{miele2025flow}), and (3) leave on a final streamline again close to the downstream separating streamline ($\eta_f\approx \eta^{sep}$). To good approximation, the important "dive" phase is described by the wall expansion model \eqref{eq wall xpnsn}, see the inset of figure~\ref{fig trajectory}(b). Noting $v_{p\perp}=a_p d\Delta/dt$, and dividing by $d\eta/dt$ along the trajectory, we write
\begin{equation} \label{eqdeltaeta}
        \frac{d(\log\Delta)}{d\eta}=1.6147\frac{a_p\kappa}{\frac{d\eta}{dt}}\,,
    \end{equation} 
an equation we want to integrate from a starting point at the end of phase (1) $(\Delta_i,\eta_i)$ to an end point at the start of phase (3) $(\Delta_f,\eta_f)$. As particle displacements are strongly dominated by the (2) phase, the results are insensitive to the choice of $\Delta_i$ and $\Delta_f$, as long as both are $\ll 1$. In particular, we can choose $\Delta_i=\Delta_f=\Delta^*$ and thus require 
\begin{equation} 
        \int_{\eta_i}^{\eta_f} 1.6147 \frac{a_p\kappa}{\frac{d\eta} {dt}} d\eta = 0 \,.
\end{equation} 
To make analytical progress, we observe that $d\eta/dt$ along the trajectory is the rate of change in the $\eta$ direction of a particle moving at a distance of $\approx a_p$ from the obstacle surface, which translates to the wall-parallel particle velocity $v_{p \parallel}$, and further to the $\eta$-component of the background velocity $u_\eta$ as defined in section \S \ref{sec result}, 
\begin{multline}\label{eqetadotvpar}
\frac{d\eta}{dt} = g(\xi_{a_p},\eta) v_{p \parallel} = g(\xi_{a_p},\eta) (1-f(\Delta)) u_{\parallel}\\= g(\xi_{a_p},\eta) (1-f(\Delta)) u_{\eta}(\xi_{a_p},\eta)=-g^2(\xi_{a_p},\eta) (1-f(\Delta))\partial_\xi\psi(\xi_{a_p},\eta)\,.
 \end{multline} 
These equations, accurate to relative order ${\cal O}(a_p^2)$, contain the scale factor of the elliptic coordinate system $g =g(\xi,\eta)$ and the $\xi$-derivative of the reference streamfunction $\psi$ (cf. sections \S\ref{sec.symmetry breaking} and \S\ref{sec result}) both evaluated at $\xi_{a_p}(\eta) = \xi_0 + a_p g(\xi_0,\eta)$, representing points a distance $a_p$ from the obstacle wall.  
The factor $f(\Delta)$ from \eqref{eq vpparallelFull} depends very weakly on $\Delta$ over the range of interest here, so that we replace it by a constant average value $\tilde{f}={\cal O}(1)$. The choice of  this constant $\bar{f}$ is irrelevant for solving \eqref{eqdeltaeta}. 
Combining the constants into $\mathcal{S}=-\frac{1.6147}{1-\tilde{f}}$, the integrand of \eqref{eqdeltaeta} becomes
\begin{equation} \label{eq phi}
       \phi(\eta)=\mathcal{S}\frac{a_p\kappa}{g^2 \partial_\xi\psi}\,,
    \end{equation}
which is now entirely defined by the background flow, and is a function of $\eta$ only. Its  behavior is discussed in more detail in appendix \ref{appen Deltapsimax}.  

Introducing for convenience the indefinite integral
$I(\eta)\equiv \int^\eta\phi(\tilde{\eta})d\tilde{\eta}$\,,
the condition \eqref{eqdeltaeta} becomes
    \begin{equation} \label{eq etafinal etaini}
        I(\eta_f)=I(\eta_i)\,,
    \end{equation}
relating valid pairs of initial and final $\eta$ coordinates. These translate into initial and final stream function values 
\begin{equation}\label{eq psistar}
\psi_i^*=\psi(\Delta=\Delta^*,\eta=\eta_i)\,,
\end{equation}
\begin{equation} \label{eq psfstar}
    \psi_f^*=\psi(\Delta=\Delta^*,\eta=\eta_f)\,.
\end{equation} 
The results of this calculation are insensitive to the precise value of $\Delta^*$ as long as it is $\ll 1$ (see appendix \ref{appen modeling} for a discussion). Figure~\ref{fig deflection}(c) 
uses $\Delta^*=0.05$ and shows good agreement with the small-$\psi_i$ trajectory data from figure ~\ref{fig deflection}(a). It is remarkable that this small-$\psi_i$ theory yields not only an asymptote for $\Delta\psi\to 0$, but captures the position of $\Delta\psi_{max}$ accurately.

\subsubsection{Scaling laws for $\Delta\psi_{max}$}\label{scaling}
The analytical approach in \S\ref{analytical} is successful, but still evaluates the necessary functions and integrals numerically. In order to understand systematically how maximum deflection depends on the parameters of the particle-obstacle encounter, we employ further simplifications.  
First, we replace $\psi$ by $\psi_B$ in the definition of $\Delta\psi$ (cf.\  \eqref{eq psiB}
), again using $\xi=\xi_{a_p}$, and furthermore expand $\psi_B$ in small $\xi-\xi_0$ (these approximations are consistent with $a_p\ll 1$).

To leading order in $a_p$, the simplified background stream function reads
\begin{equation}\label{eq streamfunction expansion in xi0}
    \hat{\psi}_{B}(\eta)=(\sin\eta\cos\alpha-\beta\cos\eta\sin\alpha)a_p^2g^2(\xi_0,\eta)\,.
\end{equation}

This stream function expression is used to evaluate initial and final values $\hat{\psi}_{i,f}=\hat{\psi}_B(\eta_{i,f}$) for given angular arguments, and also to obtain simplified versions of the functions $\phi\to\hat{\phi}$ from \eqref{eq phi} and $I\to \hat{I}$ resulting in an algebraically simplified analog of \eqref{eq etafinal etaini},
    \begin{equation} \label{eq ihat}
        \hat{I}(\eta_f)=\hat{I}(\eta_i)\,.
    \end{equation}

To directly compute the maximum of $\Delta\psi(\hat{\psi}_i)$, we need to derive a second equation. The monotonicity of
$I(\eta)$ and $\psi(\eta)$ around $\Delta\psi_{max}$ (see appendix \ref{appen Deltapsimax}) allows to first write the maximum condition as
\begin{equation} \label{eq max delfection}
    \frac{\partial\Delta\psi}{\partial \hat{I}}=0\,,
\end{equation}
and further,  defining $\zeta(\eta)\equiv \partial_{\eta}\hat{\psi}_B/\hat{\phi}$, we conclude that \eqref{eq max delfection} implies
\begin{equation}\label{eq zeta}
\zeta(\eta_f)=\zeta(\eta_i)\,.
\end{equation}
Solving \eqref{eq ihat} and \eqref{eq zeta} simultaneously yields a pair of values $(\eta_i,\eta_f)=(\eta_{i,max},\eta_{f,max})$ that determine $\Delta\psi_{max}$ from $\psi_{i,max}=\hat{\psi}_B(\eta_{i,max})$ and $\psi_{f,max}=\hat{\psi}_B(\eta_{f,max})$. To make analytical progress, it is crucial to acknowledge that the upstream and downstream angular positions $\eta_i$ and $\eta_f$ are not equidistant from the separation streamlines, but that in writing $\eta_i=\pi+\eta^{sep}-\Delta\eta_i$ and  $\eta_f=\eta^{sep}+\Delta\eta_f$, the angular deviations have a leading-order asymmetry $\delta = \Delta\eta_i-\Delta\eta_f$ (see appendix~\ref{appen Deltapsimax}), which vanishes as $\alpha\to 0$. Consistent expansion in small $\alpha$ or equivalently in small $\sin 2\alpha$ yields the leading-order expressions $\delta=\delta_1 \sin 2\alpha$ and $\eta^{sep} = (\beta/2)\sin 2\alpha$. Substituting into \eqref{eq ihat}, \eqref{eq zeta} and further expansion in the small quantities $a_p$ and $\Delta\eta_f$ results in a system of two equations that can be solved for $\delta_1$ and $\Delta\eta_f$ as detailed in appendix~\ref{appen Deltapsimax}, yielding
\begin{equation}\label{eq DeltaEtaf full}
    \Delta\eta_f=\frac{a_p\beta^2[(2-\beta^2) E(1 - \frac{1}{\beta^2})- 
        K(1 - \frac{1}{\beta^2})]}{(1-\beta^2)(6a_p+\beta^2)} \,,
\end{equation}
\begin{equation}\label{eq delta1 full}
    \delta_1=\frac{6a_p^3\beta[(2-\beta^2) E(1 - \frac{1}{\beta^2})- 
        K(1 - \frac{1}{\beta^2})]^2}{(1-\beta^2)(6a_p+\beta^2)^2} \,.
\end{equation}
Here $K$ and $E$ are the complete elliptic integrals of the first kind and second kind, respectively. 

\begin{figure}
  \centering
  \begin{subfigure}[t]{\textwidth}
    \centering
    \includegraphics[width=\textwidth]{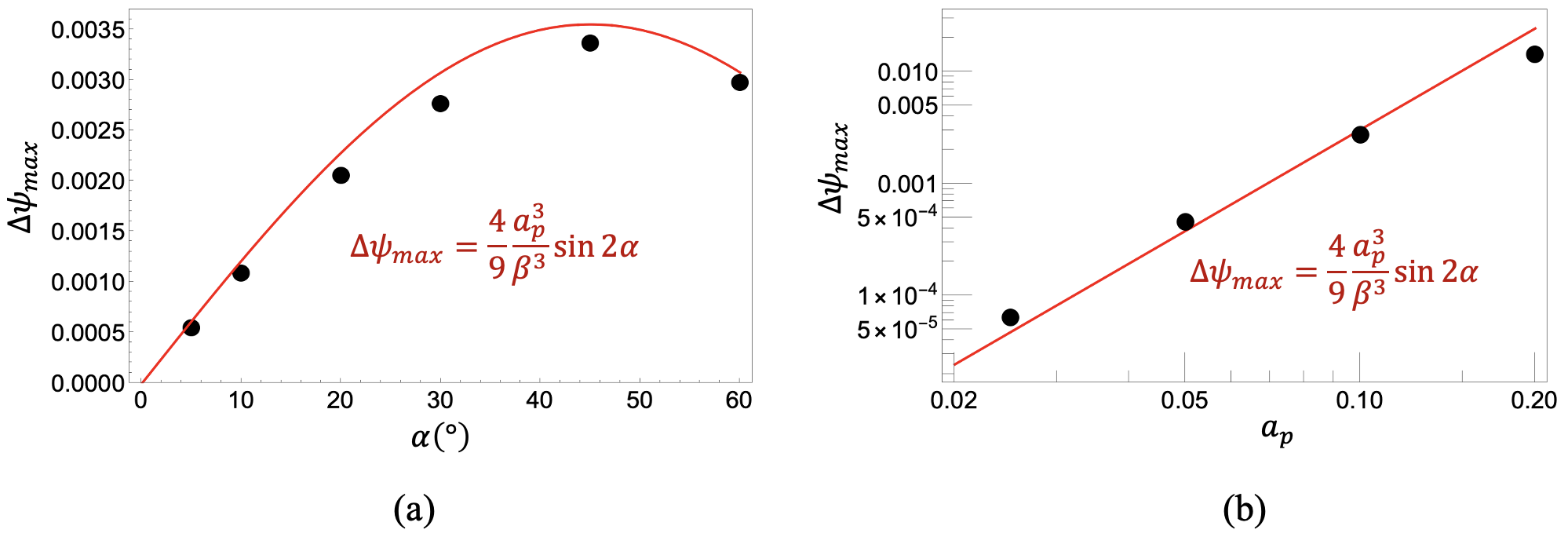} 
  \end{subfigure}

  \vspace{1em}

 \begin{subfigure}[t]{0.5\textwidth}
    \centering
    \includegraphics[width=\textwidth]{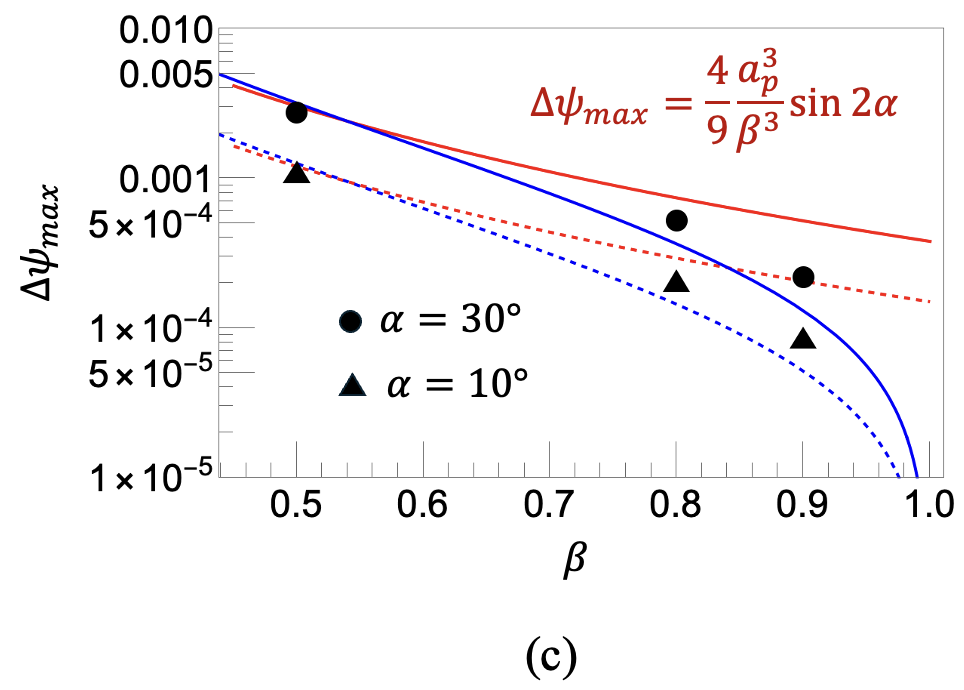}
  \end{subfigure}
  
  \caption{Scaling of $\Delta\psi_{max}$ with (a) flow angle $\alpha$ $(a_p=0.1,\beta=0.5)$, (b) particle size $a_p$ $(\alpha=30\degree,\beta=0.5)$, and (c) aspect ratio $\beta$ $(a_p=0.1,\alpha=10\degree$ and $30\degree$). The red lines are obtained from the analytical scaling theory \eqref{eq scaling psiB} for $\beta\ll 1$. In (c), 
  the small-$\beta$ theory is successful even for $\beta=0.5$, but to capture displacements for near-spherical obstacles ($\beta\lesssim 1$), the complete theory of 
  \eqref{eq DeltaEtaf full} -- \eqref{eq Deltapsimax full in delta1 Deltaetaf} is needed (blue lines). 
  }\label{fig scaling}
\end{figure}

The same consistent expansions in $\hat{\psi}_B$ eventually provide an analytical expression for $\Delta\psi_{max}$ in terms of $\delta_1$ and $\Delta\eta_f$, namely
\begin{equation}\label{eq Deltapsimax full in delta1 Deltaetaf}
    \Delta\psi_{max}=a_p^2 \left[\frac{\delta_1}{\beta^2}+\frac{2(1-\beta^2)}{\beta^3}\Delta\eta_f^2+\delta_1\frac{2(-6+5\beta^2)}{2\beta^4}\Delta\eta_f^2+O(\Delta\eta_f^3)\right]\sin{2\alpha}\,.
\end{equation}
Plugging \eqref{eq DeltaEtaf full} and \eqref{eq delta1 full} into \eqref{eq Deltapsimax full in delta1 Deltaetaf} allows for evaluation at arbitrary $\beta$ values. For $\beta\ll 1$, this expression has a much simpler limit,
\begin{equation}\label{eq scaling psiB}
   \Delta\psi_{max}= \frac{4}{9}\frac{a_p^3}{\beta^3}\sin2\alpha\qquad (\beta\ll 1)\,.
\end{equation}
Remarkably, we find that for our preferred study case $\beta=1/2$ the small-$\beta$  solution \eqref{eq scaling psiB} still reproduces trajectory data very accurately, as demonstrated by the red curves in figure~\ref{fig scaling}.
Figure~\ref{fig scaling}(a), (b) shows that
\eqref{eq scaling psiB} compares very well
with direct trajectory calculations varying flow angle of attack $\alpha$ and particle size $a_p$. The expansions behind the analytical expression maintain good accuracy beyond their strict limits: The leading-order angle dependence $\sin 2\alpha$ appears valid for all angles, and is indeed the expected dependence from the elliptic symmetry of the obstacle geometry. Particle sizes as large as $a_p=0.2$ still yield good agreement. Figure~\ref{fig scaling}(c) does find that approximating displacements around near-circular obstacles requires the full scaling theory \eqref{eq DeltaEtaf full} -- \eqref{eq Deltapsimax full in delta1 Deltaetaf}, but the displacement effect is also dramatically smaller as $\beta\to 1$ and the obstacle fore-aft symmetry vanishes. The surprisingly strong $\beta^{-3}$ dependence in \eqref{eq scaling psiB}, and its accuracy for $\beta=1/2$, suggests 
that employing even slightly more eccentric obstacles could strongly enhance particle deflection. For much smaller $\beta$, however, effects of obstacle surface curvature  on the hydrodynamic wall corrections (equation \eqref{eq rcurv} in appendix~\ref{appen flat wall}) will likely become important and modify the results. 

\subsubsection{Particle size separation}\label{separation}
We now compare the impact on particle separation by size due to the hydrodynamic effects modeled here to that inferred from the detailed modeling of short-range roughness effects \citep{frechette2009directional}. The present work treats interaction of a single particle and obstacle, and thus cannot provide critical particle sizes for crossing separating streamlines between multiple obstacles in a DLD array, but we can ask by how much the particle-obstacle interaction makes the trajectories of two particles of different sizes deviate from each other, setting up further downstream separation. We choose typical dimensional microparticle radii $a_{p1}=4\mu$m and $a_{p2}=8\mu$m, interacting with obstacles of circular cross section with $r=32\mu$m. In \citet{frechette2009directional}, the obstacles are themselves spherical and the interaction depends weakly on the  symmetry-breaking surface roughness -- we assume an experimentally relevant scale of $\sim 100$\,nm \citep{smart1989measurement,yang2007correction,hulagu2024towards}. The results in \citet{frechette2009directional} then allow for an evaluation of the displacements of the two particles when they are initially on the same streamline. Among all initial conditions, the maximum expected difference for the example parameters above is $\Delta n_{12}\approx 1.4\mu$m perpendicular to the uniform flow direction. Evaluating by comparison the hydrodynamic effects of encountering an elliptic cylinder, we choose scales of $a=40\mu$m and $b=20\mu$m (resulting in nearly the same cross-sectional area as the circular obstacle), and a flow angle of attack $\alpha=30\degree$. At a distance of $\sim a$ behind the obstacle, we find a maximal displacement difference of the same particle sizes normal to the far-field flow of $\Delta n_{12}\approx 0.9\mu$m. Thus, the short-range roughness interactions for symmetric obstacles and the hydrodynamic effects modeled here for symmetry-breaking obstacles have a comparable effect on particle separation by size and should both be taken into account when modeling DLD displacement -- we expect significant quantitative changes when obstacle cross sections are made asymmetric. 


\subsection{Closest approach and sticking} \label{sec closest approach}

Our formalism finds the strongest net deflections of particles when the trajectories follow the obstacle surface very closely ("dives", \cite{miele2025flow}). Necessarily, the minimum gap $\Delta_{min}$ on such trajectories becomes very small, and with typical microfluidic scales of obstacle and particle sizes, $\Delta_{min}$ can easily translate to sub-micrometer distances. In the presence of short-range attractive interactions between the surfaces (London forces, van der Waals attraction, etc.) this can lead to sticking (capture) of the particles, an effect important in fouling and cleaning \citep{rajendran2025recent,gul2021fouling,kumar2015fouling}, porous-media filtration \citep{miele2025flow,mays2005hydrodynamic}, or elimination of pathogens \citep{nuritdinov2025experimental,uttam2023hypothetical,sande2020new}.   
Particles used for size-based sorting, selecting, and trapping applications are often a few $\mu$m in size, so that a typical $\Delta_{min}=O(10^{-3})$ corresponds to a few nanometer gap, easily close enough for short-range attractions to be important. 



\begin{figure}
    \centering
\includegraphics[width=\textwidth]{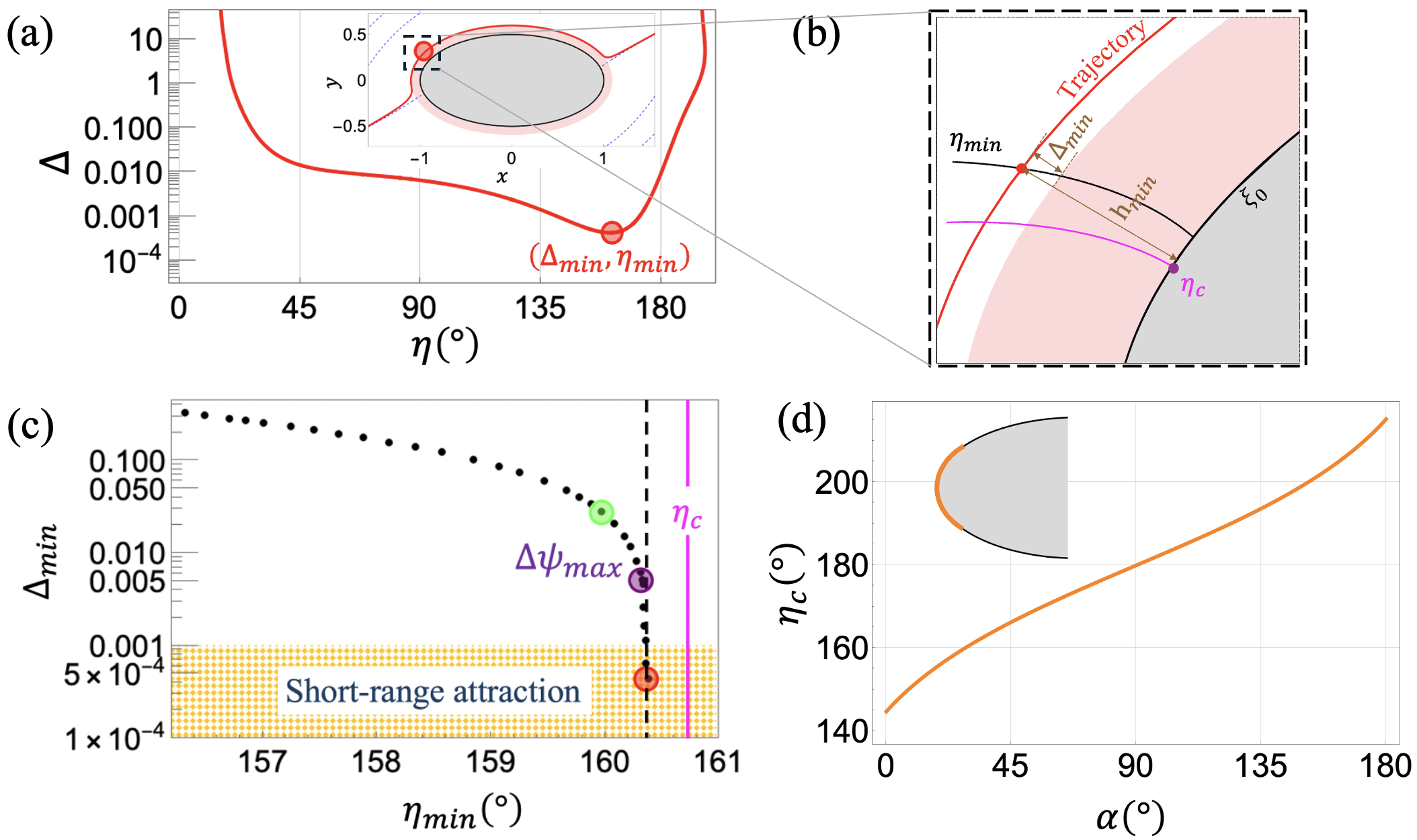}    \caption{(a) Gap size $\Delta$ as a function of $\eta$ along a "dive" trajectory creeping around the obstacle (inset). 
(b) Close-up sketch near $\eta_c$ demonstrating that the particle center angular coordinate at closest approach $\eta_{min}$ cannot coincide with $\eta_c$ if $a_p>0$.  
(c) Closest approach coordinates $(\Delta_{min}, \eta_{min})$ for different trajectories in the $\psi_i\rightarrow0$ limit. Dashed line represents $\eta_{min}(\Delta_{min}=0)$.  While $\Delta_{min}$ varies by orders of magnitude, $\eta_{min}$ stays close to the value $\eta_c$ where flow curvature $\kappa$ vanishes.
$\Delta_{min}=O(10^{-3})$ represents a few nanometer of particle to surface wall gap for a typical microparticle. Short-range inter-molecular forces activate in this range and restrict particle movement. 
(d) Variation of $\eta_c$ with $\alpha$ for $\beta=1/2$, showing that particle sticking is always most likely near the major-axis tip of the ellipse.}\label{fig minapproach}
\end{figure}

Figure~\ref{fig minapproach}(a) plots $\Delta(\eta)$ along a dive trajectory, showing a well-defined closest approach point $(\Delta_{min},\eta_{min})$. Changing $\psi_i$, figure~\ref{fig minapproach}(b) shows that $\Delta_{min}$ is very sensitive to initial conditions, but $\eta_{min}$ is not. In fact, it follows from the wall-expansion approximation \eqref{eq wall xpnsn} that the closest approach  is always determined by the point on the wall where $\kappa(\eta_c)=0$, a function of background flow only. However, for finite particle size the angular position $\eta_{min}$ of the particle where the dynamics is governed by $\eta_c$ is slightly different, as illustrated in figure~\ref{fig minapproach}(c). It is easy to show that
\begin{equation}\label{eq etamin etac connection}
    \eta_c-\eta_{min}\sim a_p^2    
\end{equation}
to leading order, so that even as $\psi_i\to 0$, a finite difference between $\eta_{min}$ and $\eta_c$ remains. For $a_p\ll 1$, practically relevant trajectories nevertheless have a very well-defined point of closest approach, to within a few degrees of $\eta_c$ (figure~\ref{fig minapproach}(b)), and thus a well-defined location where sticking is most likely follows directly from the background flow wall curvature $\kappa$. The angular position $\eta_c$ changes with the flow angle of attack $\alpha$ but not very widely. Figure~\ref{fig minapproach}(d) demonstrates that this point is located near the major-axis pole of the elliptic cylinder for all $\alpha$.


In the case of internal Stokes flow confined by flat walls \citep{liu2025principles}, the approach to minimum gap distance is well described by a single exponential, as predicted by theory \citep{brady1988stokesian,claeys1989lubrication,claeys1993suspensions}. In the present case, the strong  variation in $\kappa(\eta)$ with obstacle topography precludes such a simple behavior. The direct comparison between our full (variable expansion) formalism and a pure wall expansion calculation in the inset of figure~\ref{fig trajectory}(c) confirms, however, that the decrease of $\Delta$ over orders of magnitude in proximity to $\eta_c$ is governed by
$\kappa$. The rapid approach in the presence of a variety of short-range forces will reliably lead to sticking at this predefined location. 
\citet{pradel2024role} in their latest research on particle accumulation in transport flow through porous media (similar to the asymmetric set-up of obstacles in a DLD device), emphasize the role of hydrodynamics to arrest a microparticle on the surface of a pore but did not establish any concrete mechanism on defining specific location of single particle capture. In this context, our mechanism offers valuable strategies for microfluidic filtration and particle capture in transport flow through porous media \citep{spielman1977particle,miele2025flow,pradel2024role}.

\section{Conclusions} \label{sec conclusions}
The goal of our study has been to isolate the purely hydrodynamic displacement effect on a single force-free spherical particle encountering a cylindrical obstacle interface in zero-inertia transport flow. We show that one such encounter can indeed have a net displacement effect on particles when the geometry of the flow and obstacle break fore-aft symmetry. An exemplary case is an elliptic obstacle placed in a uniform flow with non-trivial angle of attack. Somewhat non-intuitively, there is a particular initial condition for given parameters for which the net displacement is maximal. For the practical case of particles much smaller than the obstacle ($a_p\ll 1$), this initial condition is very close to the separating streamline associated with the obstacle, and the maximum displacement (in terms of changes of stream function) is proportional to $a_p^3$ for eccentric obstacles. The effect is thus small, but also strongly dependent on particle size. A quantitative comparison with the effects of short range roughness modeling shows comparable influence of obstacle symmetry breaking on the ability to separate microparticles by size. Thus, obstacle shape should be taken into account in set-ups pursuing such goals. 

 We find that the hydrodynamic displacement effects can be maximized by choosing a flow angle $\alpha = \pi/4$ and by decreasing the aspect ratio $\beta$ of the ellipse, though corrections from obstacle surface curvature must be taken into account for very eccentric ellipses. These results do not contradict time reversibility: If time is reversed, all hydrodynamic forces also change sign together with the background flow, and particle paths are traced backwards. 



While the present work puts the elementary hydrodynamic effects behind Deterministic Lateral Displacement on a firm footing, the practical DLD set-ups consist of many circular cylinders (pillars) arranged in an array under an inclination angle to the flow direction. One could approximate the effects on the flow of groups (e.g. two) of these circular cylinders by those of an effective elliptic cylinder, as the geometric motivation for symmetry breaking remains the same. We remark here that an analogous treatment of deflection from two circular cylinders should be possible, as flow solutions and trajectory geometry can be described in bipolar coordinates similar to the present-case elliptic coordinate system. Separating streamlines connecting two obstacles then become crucial boundaries that particles of different sizes do or do not cross, leading to DLD-like separation of particle paths by row-shifting, which has not been a subject of the present study.



In a microfluidic application with many particles, the present single-particle formalism is implicitly applicable to dilute  concentrations. An important extension of this approach would be to incorporate particle-particle interactions, allowing for the assessment of non-inertial effects in particle-laden flows, which are crucial for many practical applications \citep{guha2008transport}. Another valuable generalization of the current approach is to treat non-spherical particles, whose additional degrees of freedom allow for a wider range of qualitative trajectory behaviors \citep{yerasi2022spirographic,li2024dynamics,liu2025particlethesis}. In all cases, taking symmetry-breaking hydrodynamic interactions with boundaries into account will add a previously overlooked component to particle manipulation in any viscous flow situation.

\begin{acknowledgments}
{\bf Acknowledgments:} The authors acknowledge valuable and inspiring conversations with John Brady, Blaise Delmotte,  Camille Duprat, Anke Lindner, Bhargav Rallabandi, and Howard Stone.\\ Declaration of Interests: The authors report no conflict of interest.
\end{acknowledgments}


\appendix

\section{Wall-parallel correction factor $f(\Delta)$}\label{appen wall-parallel}
As developed in \citep{liu2025principles} by systematic asymptotic matching, the wall-parallel velocity correction factor $f(\Delta)$ takes the following form:
\begin{equation} \label{f}
f(\Delta)=1-\frac{(1+\Delta)^{4}}{0.66+3.15\Delta+5.06\Delta^2+3.73\Delta^3+\Delta^4-0.27(1+\Delta)^{4} \log\left(\frac{\Delta}{1 + \Delta}\right)}\,.
\end{equation}
Figure~\ref{fig fandfaxen}(a) illustrates the agreement with the asymptote at $\Delta\gg 1$ \citep{goldman1967slow2} as well as the logarithmic approach to the lubrication-theory limit $f\to 1$ at $\Delta\to 0$ \citep{stephen1992characterization, williams1994particle}. Equation~\eqref{f} is derived for a linear wall-parallel background velocity profile, which is asymptotically accurate for small $\Delta$, the case of primary interest in this study.

\section{Effect of the bulk Fax\'en correction}\label{appen bulk Faxen}
\begin{figure}
    \centering
\includegraphics[width=\textwidth]{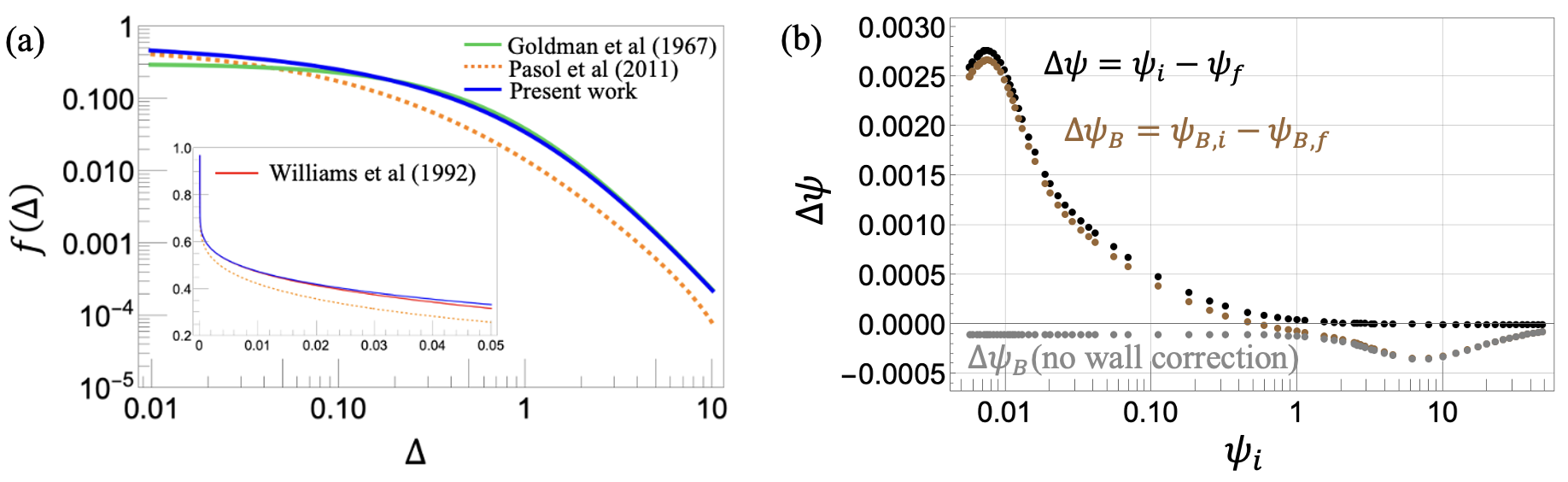}    \caption{(a) Wall parallel velocity correction factor $f(\Delta)$ as a function of non-dimensional gap $\Delta$ showing good agreement with \cite{goldman1967slow} in the far-field limit ($\Delta\gg1$) and with \cite{stephen1992characterization} (inset) in the near-field and lubrication regime ($\Delta\ll1$). (b) Comparison of displacement effects (difference between final and initial stream function values along a trajectory) with the Fax\'en contribution 
subtracted out ($\Delta\psi$) or taken into account ($\Delta\psi_B$). Fax\'en effects are
only noticeable for trajectories traveling in the far field (having larger $\psi_i$), where displacements are very small. For smaller initial conditions $\psi_i$ the displacements caused by the Fax\'en term alone (with no wall correction, gray) are insignificant compared to those caused by the wall effect. Here $(a_p,\alpha,\beta) = (0.1,30\degree,0.5)$}
    \label{fig fandfaxen}
\end{figure}

We have quantified the displacement along particle trajectories by stream function changes $\Delta\psi$ with respect to the reference flow $\psi$ from \eqref{eq faxen flow} in order to explicitly quantify the wall effects only. Even without wall effects, the presence of the Fax\'en term (the effects of bulk flow curvature) can lead to changes along the particle trajectory between the initial and final background stream function values, $\Delta\psi_B=\Delta\psi_{B,i}-\Delta\psi_{B,f}$.  
Figure~\ref{fig fandfaxen}(b) shows that this Fax\'en contribution has a noticeable effect only for initial conditions that keep the particle far from the obstacle; yet in these cases, the displacement remains extremely small. For initial conditions of smaller $\psi_i$, which yield the important displacement effects discussed in the present work, the Fax\'en effect is at least one order of magnitude smaller than the wall effects.

\section{Robustness of results against choice of modeling parameters}\label{appen modeling}
\begin{figure}
    \centering
\includegraphics[width=\textwidth]{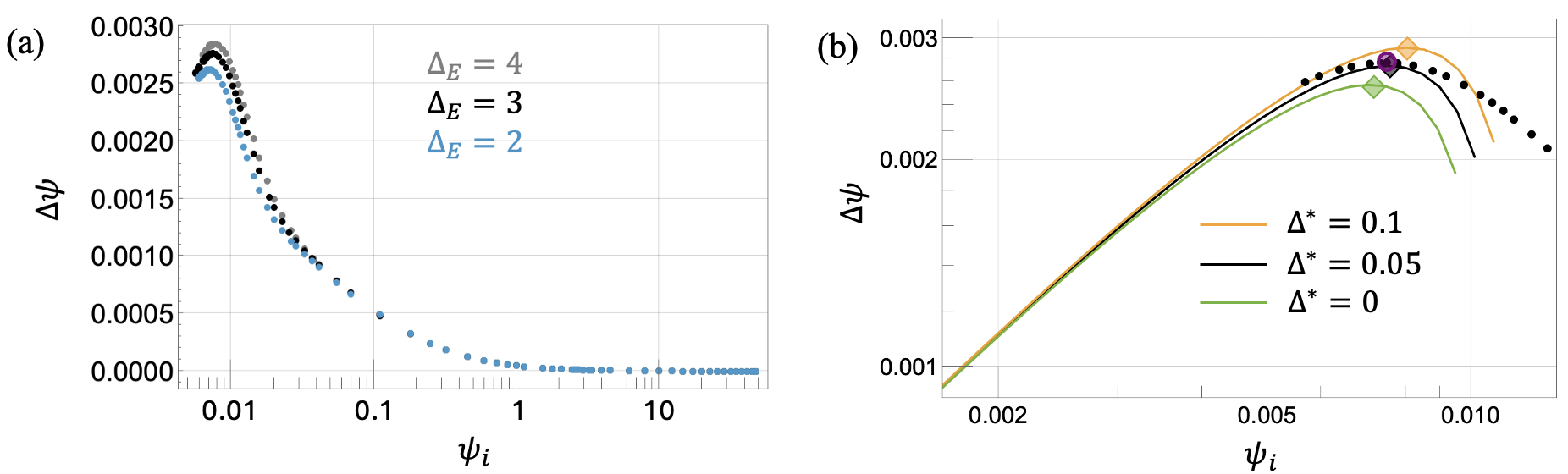}    \caption{(a) Changing the value of the modeling parameter $\Delta_E$ in the variable expansion approach by ${\cal O}(1)$ factors only weakly affects the outcome of particle displacement. (b) Results for $\Delta\psi(\psi_i)$ computed from equations \eqref{eq etafinal etaini}--\eqref{eq psfstar} with different choice of $\Delta^*$ according to the methodology discussed in \S \ref{analytical}. Black dots are the results obtained from direct numerical analysis of trajectories, cf.\ \S\ref{sec empirical}. Colored symbols identify the positions of $\Delta\psi_{max}$. All results are for $(a_p,\alpha,\beta) = (0.1,30\degree,0.5)$.
}
    \label{fig Deltastar}
\end{figure}

As mentioned in \S\ref{sec:wall normal} of the main text, the general approach to determining the 
wall-normal correction of the particle velocity at moderate to large $\Delta$ involves expanding the background velocity field around the particle center position, cf.\ \eqref{eq vpPE}, while when the particle gets very close to touching the obstacle ($\Delta\to 0$) the description asymptotes to \eqref{eq wall xpnsn}, where the velocity
is expanded around a point at the obstacle wall.
In order to smoothly transition from one description to the other, we introduce the variable expansion formalism \eqref{eq vpVE}, moving the expansion point continuously with the value of $\Delta$, an approach already utilized in our recent study on particle dynamics in internal Stokes flows \citep{liu2025principles}. 
We tested the robustness of this approach in the present case against changing the functional form of the expansion point dependence \eqref{eq xE} and found that variations are small as long as the overall smoothness of the transition is preserved. In figure~\ref{fig Deltastar}(a), we show small quantitative differences upon changing the transition parameter $\Delta_E={\cal O}(1)$, demonstrating that the displacement effect as a whole, and its maximum magnitude, are insensitive to such modeling changes.

As part of establishing the analytical model of $\Delta\psi$ from \eqref{eq etafinal etaini}, we need to specify a gap value $\Delta^*$ to compute $\psi^*_i$ and $\psi_f^*$ from \eqref{eq psistar} and \eqref{eq psfstar}, respectively, as described in section  \S \ref{analytical}. We see from figure ~\ref{fig Deltastar}(b) that the choice of $\Delta^*$ in evaluating $\psi^*_i$ and $\psi^*_f$ only weakly affects  the ultimate displacement $\Delta \psi=\psi^*_i-
\psi_f^*$, as long as $\Delta^*\ll 1$. In particular, the existence and position of a maximum in $\Delta\psi$ are robust against that choice. This motivated the simplified analytical treatment in \S\ref{scaling}, where most quantities are evaluated at a distance of $a_p$ from the obstacle, i.e., at $\Delta=0$.  

\section{Effect of wall curvature}\label{appen flat wall}
The results presented in the main text are obtained from wall corrections assuming a flat wall \citep{rallabandi2017hydrodynamic}, which is self-consistent in the limit of small gaps, $\Delta<1$. While it is intuitive that the majority of the wall interaction effects happen under this condition, one has to verify whether interactions at larger $\Delta$ accumulate to appreciable deflections and, if so, whether obstacle curvature must then be taken into account.

\begin{figure}
    \centering
\includegraphics[width=\textwidth]{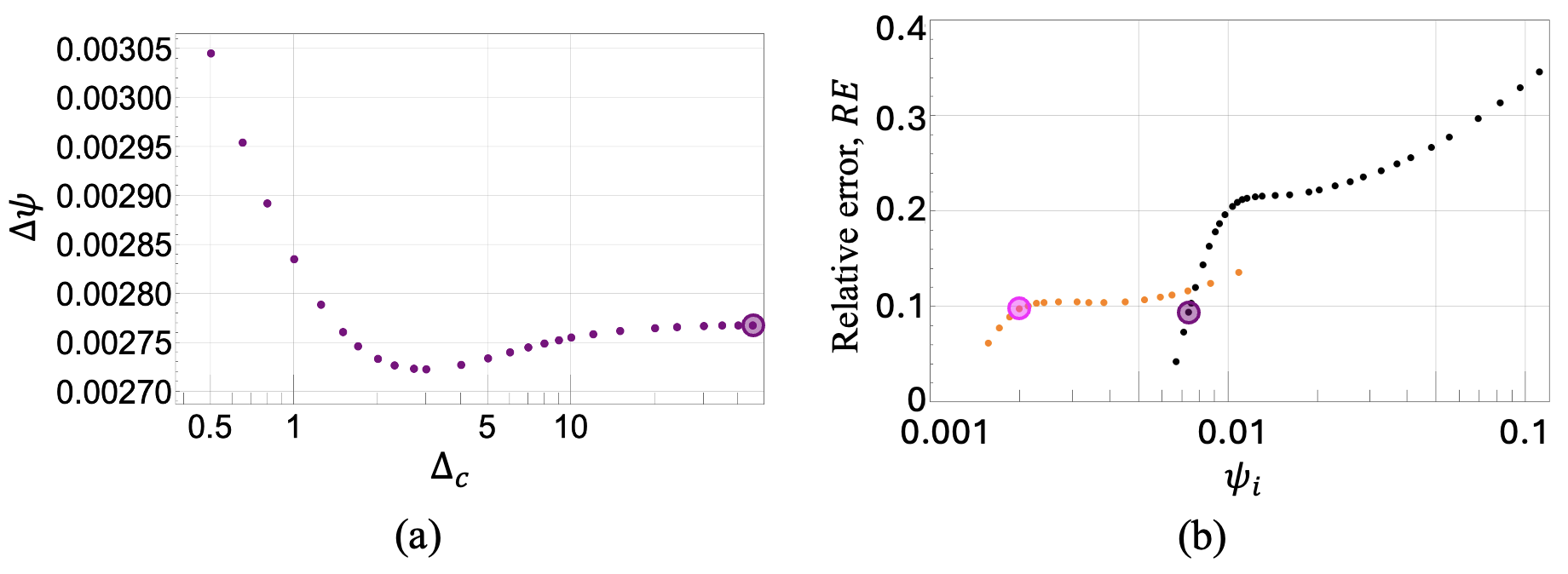}    \caption{(a) Computation of $\Delta\psi$ for $\psi_i$ corresponding to the "purple" trajectory $(\Delta\psi_{max})$ in figure~\ref{fig deflection} with wall effect $W_{\perp}$ turned on for $\Delta<\Delta_c$; for results in the main text $W_{\perp}$ was turned on everywhere (purple circle on the right). The wall effects accumulated in the far-field have negligible effect on the displacement, as choosing $\Delta_c\gtrsim1$ does not affect the final outcome appreciably while a very late activation of $W_{\perp}$  misses some important effects. Therefore, we recomputed $\Delta\psi$ for trajectories with $\Delta_{min}\lesssim 1$ taking wall curvature into account (b). 
Plotted is the relative error $RE=(\Delta\psi_{curved\, wall}-\Delta\psi_{flat\, wall})/\Delta\psi_{curved\,wall}$ for two particle sizes, $a_p=0.1$ in black and $a_p=0.05$ in orange (magenta data corresponds to $\Delta\psi_{max}$ for $a_p=0.05$). Ignoring wall curvature when the wall effect is important $(\Delta_{min}\leq\Delta_c)$  underestimates the deflection, but RE remains small $(<10\%)$ in computing the maximum deflection. The control parameters are $(\alpha,\beta) = (30\degree,0.5)$.}
    \label{fig flat wall}
\end{figure}

To address the first point, we present in figure~\ref{fig flat wall}(a) computations of $\Delta\psi$ along the "purple" type trajectory in figure~\ref{fig deflection} (exhibiting maximum displacement) where we only activate the wall effects for $\Delta<\Delta_c$, setting $W_\perp$ to zero for larger distances. The data presented in the main text are for $\Delta_c\sim 50$ (wall effects are activated everywhere, larger purple circle). The figure shows that wall effects are negligible for $\Delta_c\gtrsim 1$, confirming that (i) the vast majority of wall effect displacement happens at small $\Delta$, and (ii) the modeling of a distant elliptical obstacle at large distance as a flat wall does not introduce significant errors.

For the parts of the trajectory where $\Delta\leq\Delta_c$, is it important to model the finite curvature of the obstacle wall? We recomputed $\Delta\psi$ for the trajectories that come close enough to the obstacle to make the wall effect  important $(\Delta_{min}\leq\Delta_c)$ using the curved-wall formalism developed for spherical obstacles in \cite{rallabandi2017hydrodynamic} by evaluating the radius of curvature $R$ of the wall point closest to the particle by 
\begin{equation}
R = \frac{1}{\beta}\left(\beta ^2 \cos^2{\eta}+\sin^2{\eta}\right)^{3/2} \,.
    \label{eq rcurv}
\end{equation}
Figure ~\ref{fig flat wall}(b) shows that the relative error in $\Delta\psi$ using the flat-wall formalism sharply drops for particles passing the obstacle in closer proximity (in particular for trajectories near maximum deflection), while it becomes small throughout for smaller particles. Note that this computation likely overestimates  the influence of curvature, as \cite{rallabandi2017hydrodynamic} describe the effect of a nearby spherical obstacle (two radii of curvature); qualitatively, our computation indicates that the displacements are slightly enhanced by curvature. However, because the curvature formalism is computationally expensive and does not introduce large errors, we use the flat-wall formalism for all computations presented in the main text.  
 


\begin{figure}
    \centering
\includegraphics[width=0.5\textwidth]{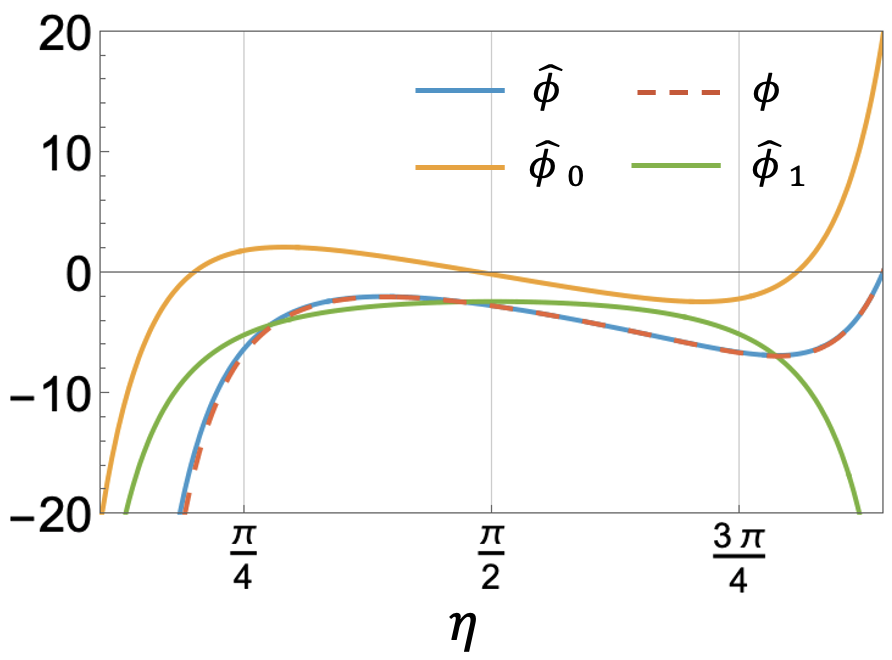}    \caption{The analytically integrable function $\hat{\phi}$ as introduced in section \S\ref{scaling} is in excellent agreement with the $\phi$ function \eqref{eq phi} developed in section \S\ref{analytical}.   We expand $\hat{\phi}$ in $\sin{2\alpha}$ as $\hat{\phi}=\hat{\phi}_0+\hat{\phi}_1\sin{2\alpha+O(\sin^2{2\alpha})}$ where the leading order term $\hat{\phi}_0$ is antisymmetric but the first order term $\hat{\phi}_1$ is symmetric around $\eta=\pi/2$. Here $(a_p,\alpha,\beta)\equiv(0.1,30\degree,0.5)$.}
    \label{fig phi}
\end{figure}

\section{Scaling of $\Delta\psi_{max}$ with parameters}\label{appen Deltapsimax}
The central quantity on which analytical computation of displacement hinges is the function $\phi(\eta)$ from \eqref{eq phi} depicted in figure~\ref{fig phi}. We know that $\phi$ must be symmetric with respect to $\eta=\pi/2$ when $\alpha=0$, and proceed to expand it for small $\alpha$. Consistent to leading order, we choose $\sin 2\alpha$ as our expansion parameter, which conforms with the expected behavior at larger $\alpha$. Expanding as far as $\phi=\phi_0+\phi_1\sin 2\alpha$, equation~\eqref{eq etafinal etaini} reads
\begin{equation}\label{eqinitial}
    \int_{\eta_f}^{\eta_i}(\phi_0+\phi_1\sin{2\alpha}) d\eta=0\,.
\end{equation}
Making use of the simplified background stream function \eqref{eq streamfunction expansion in xi0} and consistent evaluation of the terms in \eqref{eq phi} at $a_p$ distance to the obstacle ($\Delta\to 0$), 
we obtain analytically integrable versions of the two leading-order functions in \eqref{eqinitial}, namely 
\begin{multline}
    \hat\phi_0=\frac{(1-\beta ^2)\left(3 \beta ^2-1+\left(1-\beta ^2\right) \cos {2 \eta}\right)}{
   \left(1-\left(1-\beta ^2\right) \cos {2\eta}+\beta ^2\right)^{5/2}\tan {\eta}}\\\left(\cos {2 \eta}-\cosh 
   \left(\frac{2\sqrt{2} a_p}{\sqrt{1-\left(1-\beta^2\right) \cos {2 \eta}+\beta^2}}+2\tanh
   ^{-1}\beta\right)\right)
\end{multline}

and
\begin{multline}
    \hat\phi_1=\frac{\beta (1-\beta^2) }{(1-(1-\beta^2 )\cos {2 \eta}+\beta^2)(1-\cos{2\eta})}\\\left(\cos {2 \eta}-\cosh 
   \left(\frac{2\sqrt{2} a_p}{\sqrt{1-\left(1-\beta^2\right) \cos {2 \eta}+\beta^2}}+2\tanh
   ^{-1}\beta\right)\right)\,.
\end{multline}
As can be seen in figure ~\ref{fig phi}, $\hat\phi_0$ is odd and $\hat\phi_1$ is even with respect to $\eta=\pi/2$ (and the sum of both terms is an excellent approximation of $\phi$).
However, integration over $\hat\phi_0$ from $\eta_i=\pi+\eta^{sep}-\Delta\eta_i$ to $\eta_f=\eta^{sep}+\Delta\eta_f$ does not respect these symmetries for two reasons: $\eta^{sep}\not = 0$ and $\delta=\Delta\eta_i-\Delta\eta_f\not = 0$. Both quantities do vanish as $\alpha\to 0$; from its definition we have $\eta^{sep} =(\beta/2)\sin 2\alpha$ to leading order and we write $\delta=\delta_1 \sin 2\alpha$ with a $\delta_1$ to be determined. Applying these definitions in \eqref{eqinitial}, all non-zero terms are proportional to $\sin 2\alpha$ and we obtain an explicit equation for $\delta_1$,

\begin{equation}\label{eq delta1 from phi}
    \delta_1=\frac{\int_{\eta=\Delta\eta_f}^{\eta=\pi-\Delta\eta_f}\hat\phi_1 d\eta}{\hat\phi_0(\eta=\pi-\Delta\eta_f)}+\beta\,.
\end{equation}

Furthermore, we expand \eqref{eq delta1 from phi} in small $a_p$ and small $\Delta\eta_f$ to obtain
\begin{equation}\label{eq delta1 from phi leading}
    \begin{split}
        \delta_1=2\bigg[& a_p\frac{(2-\beta^2) E\left(1 - \frac{1}{\beta^2}\right)- 
        K\left(1 - \frac{1}{\beta^2}\right)}{\beta}\Delta\eta_f \\& -\frac{(1- \beta^2) (\beta^2 + 3 a_p)}{\beta^3}\Delta\eta_f^2+O(\Delta\eta_f^3)\bigg]+O(a_p^2)
    \end{split}
\end{equation}
with the complete elliptic integrals of first ($K$) and second ($E$) kind.

Equation~\eqref{eq delta1 from phi leading} computes $\delta$, and thus $\eta_i$, from a given $\Delta\eta_f$, and thus $\eta_f$. To isolate the values for maximum particle deflection, we start with the definition of the extremum 
\begin{equation}\label{eq max Deltapsi psi}
    \frac{\partial\Delta\psi}{\partial\psi_i}=0
\end{equation}
and using the chain rule and the simplified streamfunction $\hat{\psi}_B$ from \eqref{eq streamfunction expansion in xi0} to obtain
\begin{equation}\label{eq max Deltapsi psi 2}
  \frac{\partial\Delta\psi}{\partial\hat{\psi}_B}=\frac{\partial\Delta\psi}{\partial \hat{I}}\frac{\partial_\eta \hat{I}}{\partial_\eta\hat{\psi}_B}\,.
\end{equation}
One can find that both $\partial_\eta \hat{I}=\hat{\phi}$ and $\partial_\eta \hat{\psi}_B$ are non-zero over the range of $\eta$ of interest for evaluating the maximum, so that we can instead use
equation \eqref{eq max delfection} as the maximum condition. 

We here define $\zeta(\eta)\equiv \partial_{\eta}\hat{\psi}_B/\hat{\phi}$ (cf. section \S \ref{scaling}) and rewrite \eqref{eq zeta} as
\begin{equation}\label{eq zeta Deltaeta}
    \zeta(\eta_i=\pi+\eta^{sep}-\Delta\eta_f-\delta)=\zeta(\eta_f=\eta^{sep}+\Delta\eta_f)\,.
\end{equation}
Employing the same expansion in small $\delta$ as above, we find
\begin{equation}\label{eq delta from zeta}
\delta=\delta_1\sin{2\alpha}=\frac{\zeta(\eta=\pi+\eta^{sep}-\Delta\eta_f)-\zeta(\eta=\eta^{sep}+\Delta\eta_f)}{\partial_{\eta}\zeta|_{\eta=\pi+\eta^{sep}-\Delta\eta_f}}\,,
\end{equation}
and expansion of the RHS in small $\eta^{sep}$ again yields terms proportional to $\sin 2\alpha$.
Expanding in small $a_p$ and 
$\Delta\eta$ 
results in another leading-order expression for $\delta_1$,
\begin{equation}\label{eq delta zeta leading}
    \delta_1=\left[6a_p\frac{1-\beta^2}{\beta^3}\Delta\eta_f^2+O(\Delta\eta_f^3)\right]+O(a_p^2)
\end{equation}
Solving \eqref{eq delta1 from phi leading} and \eqref{eq delta zeta leading} simultaneously for $\delta_1$ and $\Delta\eta_f$, we obtain the expressions \eqref{eq DeltaEtaf full} and \eqref{eq delta1 full}. These eequations can be evaluated in the limit of  $\beta\ll 1$ to yield
\begin{equation}\label{eq DeltaEtaf}
    \Delta\eta_f=\frac{\beta}{3}\,,
\end{equation}
\begin{equation}\label{eq delta1}
    \delta_1=\frac{2}{3}\frac{a_p}{\beta}\,,
\end{equation}
which after insertion in
\eqref{eq Deltapsimax full in delta1 Deltaetaf}  obtains the analytical limit of $\Delta\psi_{max}$  for small $\beta$ as given in \eqref{eq scaling psiB}. Note that only the $\delta_1$ terms in \eqref{eq Deltapsimax full in delta1 Deltaetaf} contribute in this limit.





\bibliographystyle{jfm}
\bibliography{jfm-instructions}

\end{document}